\RequirePackage{ifpdf}
\documentclass[12pt,letterpaper]{article}
\usepackage{jheppub}
\usepackage{epsfig}
\usepackage[latin1]{inputenc}
\usepackage{bbm,amsfonts}
\usepackage{graphicx}
\usepackage{amssymb,amsmath}
\usepackage{fancybox}
\usepackage{verbatim}

\author[a]{Marco S. Bianchi,}
\author[b]{ Luca Griguolo,}
\author[c]{ Matias Leoni,}
\author[d]{Andrea Mauri,}
\author[d,e]{\\Silvia Penati}
\author[f]{and Domenico Seminara}
 \preprint{QMUL-PH-16-13}
\affiliation[a]{Center for Research in String Theory - School of Physics and Astronomy Queen Mary University of London, Mile End Road, London E1 4NS, UK}
\affiliation[b]{Dipartimento di Fisica e Scienze della Terra, Universit\`a di Parma and INFN Gruppo Collegato di Parma, Viale G.P. Usberti 7/A, 43100 Parma, Italy}
\affiliation[c]{Physics Department, FCEyN-UBA \& IFIBA-CONICET\ 
  Ciudad Universitaria, Pabell\'on I, 1428, Buenos Aires, Argentina }
\affiliation[d]{ Dipartimento di Fisica, Universit\`a degli studi di Milano--Bicocca, Piazza della Scienza 3, I-20126 Milano, Italy }
\affiliation[e]{INFN, Sezione di Milano--Bicocca, Piazza della Scienza 3, I-20126 Milano, Italy }
\affiliation[f]{Dipartimento di Fisica, Universit\`a di Firenze and INFN Sezione di Firenze, via G. Sansone 1, 50019 Sesto Fiorentino, Italy\\}

\emailAdd{m.s.bianchi@qmul.ac.uk}  
\emailAdd{luca.griguolo@pr.infn.it} 
\emailAdd{andrea.mauri@mi.infn.it} 
\emailAdd{leoni@df.uba.ar}
\emailAdd{silvia.penati@mib.infn.it} 
\emailAdd{seminara@fi.infn.it}

\abstract{In three dimensional ${\cal N}=4$ Chern--Simons-matter theories two independent fermionic Wilson loop operators can be defined, which preserve half of the supersymmetry charges and are cohomologically equivalent at classical level.
We compute their three-loop expectation value in a convenient color sector and prove that the degeneracy is uplifted by quantum corrections. We expand the matrix model prediction in the same regime and by comparison we conclude that the quantum 1/2 BPS Wilson loop is the average of the two operators. We provide an all-loop argument to support this claim at any order. As a by--product, we identify the localization result at three loops as a correction to the framing factor induced by matter interactions. Finally, we comment on the quantum properties of the non--1/2 BPS Wilson loop operator defined as the difference of the two fermionic ones.}

\title{The quantum 1/2 BPS Wilson loop in ${\cal N}=4$ Chern--Simons-matter theories}


\csname @addtoreset\endcsname{equation}{section}

\def\bseq{\begin{subequation}}  
\def\eseq{\end{subequation}}
\def\bsea{\begin{subeqnarray}}  
\def\esea{\end{subeqnarray}}

\def\Tilde#1{\widetilde{#1}}                    

\newcommand{\beq}{\begin{equation}}
\newcommand{\bea}{\begin{eqnarray}}
\newcommand{\eea}{\end{eqnarray}}
\newcommand{\eeq}{\end{equation}}

\newcommand {\non}{\nonumber}

\renewcommand{\a}{\alpha}
\renewcommand{\b}{\beta}

\renewcommand{\d}{\delta}
\newcommand{\pa}{\partial}
\newcommand{\g}{\gamma}
\newcommand{\G}{\Gamma}

\newcommand{\e}{\epsilon}

\renewcommand{\l}{\lambda}

\newcommand{\m}{\mu}

\newcommand{\n}{\nu}

\newcommand{\p}{\pi}

\newcommand{\s}{\sigma}

\renewcommand{\t}{\tau}

\newcommand{\be}{\begin{equation}}
\newcommand{\ee}{\end{equation}}
\newcommand{\nn}{\nonumber}

\def\Mb{\kern 2pt\mathchoice
        {
         \vbox{\hrule width10pt height 0.4pt depth 0pt
         \kern 1.2pt\hbox{\kern -2pt$\displaystyle M$}}}
        {
         \vbox{\hrule width10pt height 0.4pt depth 0pt
         \kern 1.2pt\hbox{\kern -2pt$\textstyle M$}}}
        {
\vbox{\hrule width6pt height 0.4pt depth 0pt
         \kern 1.0pt\hbox{\kern -2pt$\scriptstyle M$}}}
        {
         \vbox{\hrule width5pt height 0.4pt depth 0pt
         \kern 0.8pt\hbox{\kern -2pt$\scriptscriptstyle M$}}}}

\def\Sb{\kern 2pt\mathchoice
        {
         \vbox{\hrule width6pt height 0.4pt depth 0pt
         \kern 1.2pt\hbox{\kern -2pt$\displaystyle S$}}}
        {
         \vbox{\hrule width6pt height 0.4pt depth 0pt
         \kern 1.2pt\hbox{\kern -2pt$\textstyle S$}}}
        {
         \vbox{\hrule width3.5pt height 0.4pt depth 0pt
         \kern 1.0pt\hbox{\kern -2pt$\scriptstyle S$}}}
        {
         \vbox{\hrule width3pt height 0.4pt depth 0pt
         \kern 0.8pt\hbox{\kern -2pt$\scriptscriptstyle S$}}}}

\def\Rb{\kern 2pt\mathchoice
        {
         \vbox{\hrule width5.5pt height 0.4pt depth 0pt
         \kern 1.2pt\hbox{\kern -2.5pt$\displaystyle R$}}}
        {
         \vbox{\hrule width5.5pt height 0.4pt depth 0pt
         \kern 1.2pt\hbox{\kern -2.5pt$\textstyle R$}}}
        {
         \vbox{\hrule width3.5pt height 0.4pt depth 0pt
         \kern 1.0pt\hbox{\kern -2.2pt$\scriptstyle R$}}}
        {
         \vbox{\hrule width3pt height 0.4pt depth 0pt
         \kern 0.8pt\hbox{\kern -2.2pt$\scriptscriptstyle R$}}}}

  \def\pp{{\mathchoice
      {
          \kern 1pt%
          \raise 1pt
          \vbox{\hrule width5pt height0.4pt depth0pt
            \kern -2pt
            \hbox{\kern 2.3pt
              \vrule width0.4pt height6pt depth0pt
              }
            \kern -2pt
            \hrule width5pt height0.4pt depth0pt}%
            \kern 1pt
       }
        {
          \kern 1pt%
          \raise 1pt
          \vbox{\hrule width4.3pt height0.4pt depth0pt
            \kern -1.8pt
            \hbox{\kern 1.95pt
              \vrule width0.4pt height5.4pt depth0pt
              }
            \kern -1.8pt
            \hrule width4.3pt height0.4pt depth0pt}%
            \kern 1pt
        }
        {
          \kern 0.5pt%
          \raise 1pt
          \vbox{\hrule width4.0pt height0.3pt depth0pt
            \kern -1.9pt  
            \hbox{\kern 1.85pt
              \vrule width0.3pt height5.7pt depth0pt
              }
            \kern -1.9pt
            \hrule width4.0pt height0.3pt depth0pt}%
            \kern 0.5pt
        }
        {
          \kern 0.5pt%
          \raise 1pt
          \vbox{\hrule width3.6pt height0.3pt depth0pt
            \kern -1.5pt
            \hbox{\kern 1.65pt
              \vrule width0.3pt height4.5pt depth0pt
              }
            \kern -1.5pt
            \hrule width3.6pt height0.3pt depth0pt}%
            \kern 0.5pt
        }
    }}

  \def\mm{{\mathchoice
               {
                 \kern 1pt
           \raise 1pt    \vbox{\hrule width5pt height0.4pt depth0pt
                  \kern 2pt
                  \hrule width5pt height0.4pt depth0pt}
                 \kern 1pt}
               {
                \kern 1pt
           \raise 1pt \vbox{\hrule width4.3pt height0.4pt depth0pt
                  \kern 1.8pt
                  \hrule width4.3pt height0.4pt depth0pt}
                 \kern 1pt}
               {
                \kern 0.5pt
           \raise 1pt
                \vbox{\hrule width4.0pt height0.3pt depth0pt
                  \kern 1.9pt
                  \hrule width4.0pt height0.3pt depth0pt}
                \kern 1pt}
               {
               \kern 0.5pt
         \raise 1pt  \vbox{\hrule width3.6pt height0.3pt depth0pt
                  \kern 1.5pt
                  \hrule width3.6pt height0.3pt depth0pt}
               \kern 0.5pt}
               }}

\def\pd{{\kern0.5pt
           + \kern-5.05pt \raise5.8pt\hbox{$\textstyle.$}\kern
0.5pt}}

\def\pmd{{\kern0.5pt
          \pm \kern-5.05pt
\raise6.3pt\hbox{$\textstyle.$}\kern1.5pt}}

\def\md{{\mathchoice
   {
      {{\kern 1pt - \kern-6.2pt \raise5pt\hbox{$\textstyle.$}\kern
1pt}}}
    {
      {{\kern 1pt - \kern-6.2pt \raise5pt\hbox{$\textstyle.$}\kern
1pt}}}
    {
      {\kern0.5pt - \kern-5.05pt
\raise3.4pt\hbox{$\textstyle.$}\kern0.5pt}}
    {
      {\kern0.5pt - \kern-5.05pt
\raise3.4pt\hbox{$\textstyle.$}\kern0.5pt}}}}

\def\beq{\begin{equation}}
\def\eeq{\end{equation}}
\def\bea{\begin{eqnarray}}
\def\eea{\end{eqnarray}}
\def\Tr{\textstyle{\rm Tr}}

\def\a{\alpha}
\def\b{\beta}
\def\g{\gamma}

\def\d{\delta}
\def\e{\epsilon}

\def\l{\lambda}
\def\G{\Gamma}

\begin{document}

\maketitle

\section{Introduction}

In this paper we continue the study of 1/4 and 1/2 BPS Wilson loops in ${\cal N}=4$ Chern-Simons (CS) theories with matter, initiated in \cite{Griguolo:2015swa}.
These operators were defined in \cite{Cooke:2015ila,OWZ1, OWZ4, OWZ2, OWZ3} and we review their construction in Section \ref{generalities} along with a quick glimpse at the structure of the ${\cal N}=4$ CS models \cite{Gaiotto:2008sd,Hosomichi:2008jd}.

The interest in supersymmetric Wilson operators arises since they are amenable of an exact computation via localization, then providing observables interpolating from weak to strong coupling \cite{Pestun:2007rz}.
Their determination is usually highly constrained by supersymmetry invariance.
For the class of theories under investigation, though, a classical analysis allows to define two seemingly independent 1/2 BPS circular loops, and any arbitrary combination thereof naively provides a supersymmetric observable \cite{Cooke:2015ila}.
Such operators possess a coupling to fermions, encapsulated in a supermatrix structure, and are cohomologically equivalent to a combination of bosonic 1/4 BPS Wilson loops, in a fashion similar to the one that links 1/2 and 1/6 BPS operators \cite{Drukker:2009hy} in the ABJ(M) models \cite{ABJM, ABJ}.
The expectation value of 1/4 BPS operators can be computed via a matrix model average, which in turn allows for the exact computation of the 1/2 BPS circular Wilson loops if the aforementioned cohomological relation survives at quantum level.

At strong coupling the dual string theory description differs from the weak regime picture outlined above. In particular, the brane configuration corresponding to the 1/2 BPS operator is expected to be unique, in contrast with the existence of a whole family of observables predicted by field theoretical analysis.

In \cite{Cooke:2015ila} a solution to this tension was proposed by suggesting that only one combination of operators should be exactly 1/2 BPS at quantum level, that is the classical degeneracy of Wilson loops should be uplifted by quantum corrections. 
If this is the case, the localization prediction turns out to be relevant only for such an exactly BPS operator. However, since it is based on the cohomological relations derived at classical level, it does not shed any light on which the correct BPS combination should be.

The question of Wilson loops degeneracy and the determination of the quantum 1/2 BPS operator can instead be answered through a perturbative evaluation of the expectation values of these operators.
Such a study was initiated in \cite{Griguolo:2015swa}, where a full-blown two-loop computation was performed, which did not find any uplift of the degeneracy, thus leaving the question open. Providing a definite answer to this problem is the main purpose of this paper.

Focusing on necklace quiver ${\cal N}=4$ CS--matter theories with gauge group $U(N_0) \times U(N_1) \times \cdots U(N_{2r-1})$  we carry out this program as follows.

\vskip 5pt
\noindent
$\bullet$ In Section \ref{sec:relation}, using Feynman rules and power counting arguments together with the definition of the two seemingly independent 1/2 BPS operators, we first prove that as a consequence of the contour planarity  their perturbative expectation values coincide at any even loop order, while they are opposite at odd loops.
As a consequence, a quantum uplift of the operators, if any, has to appear at odd orders.
This explains why no degeneracy has been found so far: The operators are vanishing at one loop, therefore not allowing for any uplift, while their expectation values coincide at two loops, on general grounds. 

\vskip 5pt
\noindent
$\bullet$ We are then forced to perform a calculation at three loops, being it the first possible order where a non-vanishing and opposite contribution to the two operators may occur. A complete three-loop computation is of course daunting, but since we are just looking for a smoking gun of the quantum uplift of degeneracy, it is sufficient to focus on a particular color sector where a limited number of non--vanishing diagrams appears. Precisely, we restrict to the sector including contributions proportional to the product of three different colors, $N_{A-1} N_A N_{A+1}$. We stress that this simplification has been made possible by the fact that we work with  quiver theories with a different gauge group in each node. 

\vskip 5pt
\noindent
$\bullet$  In Section \ref{sec:MM} we first expand the matrix model at the desired perturbative order and in the selected color sector, in order to be able to compare it with the Feynman diagram computation.
We find that at third order a non--vanishing, purely imaginary correction appears. Comparing it with a perturbative calculation done at non--vanishing framing, we prove that this contribution corresponds to a loop correction to the framing factor of the Wilson loop due to interacting matter \cite{Bianchi:2016yzj}. Therefore, we expect no three-loop corrections to the expectation value of the actual 1/2 BPS operator when computed in ordinary perturbation theory at framing zero.

\vskip 5pt
\noindent
$\bullet$  In Section \ref{lifting} we finally perform the three-loop perturbative evaluation of the Wilson loops in the aforementioned regime. We find that a non-vanishing correction indeed appears, which is opposite in sign for the two operators. This proves that the degeneracy of the operators is uplifted quantum mechanically at this order. 
Moreover, since from the matrix model expansion for the 1/2 BPS operator we expect a vanishing result, we conclude that the quantum supersymmetric Wilson loop is given by the average of the two operators
\beq
W_{1/2} = \frac{W_{\psi_1} + W_{\psi_2}}{2}
\eeq
where odd orders cancel out.  We argue that this relation holds at all orders in perturbation theory.

\vskip 7pt
 
Finally, it is interesting to note that the Wilson loop operator defined by the difference $(W_{\psi_1} - W_{\psi_2})$, although non-1/2 BPS, exhibits interesting quantum properties. In fact, thanks to the relation that holds at even and odd orders in the expansion of the two original Wilson loops, this operator has a real non--vanishing expectation value given by a purely odd perturbative series. Moreover, as comes out from our explicit calculation at three loops, it seems to feature lower transcendentality.

\section{BPS Wilson loops in ${\cal N}=4$ CS--matter theories}\label{generalities}

We begin by reviewing BPS Wilson loop (WL) operators for ${\cal N}=4$ CS--matter theories introduced in \cite{Cooke:2015ila, OWZ1}. 

\begin{figure}
\centering
 \includegraphics[width=0.40\textwidth]{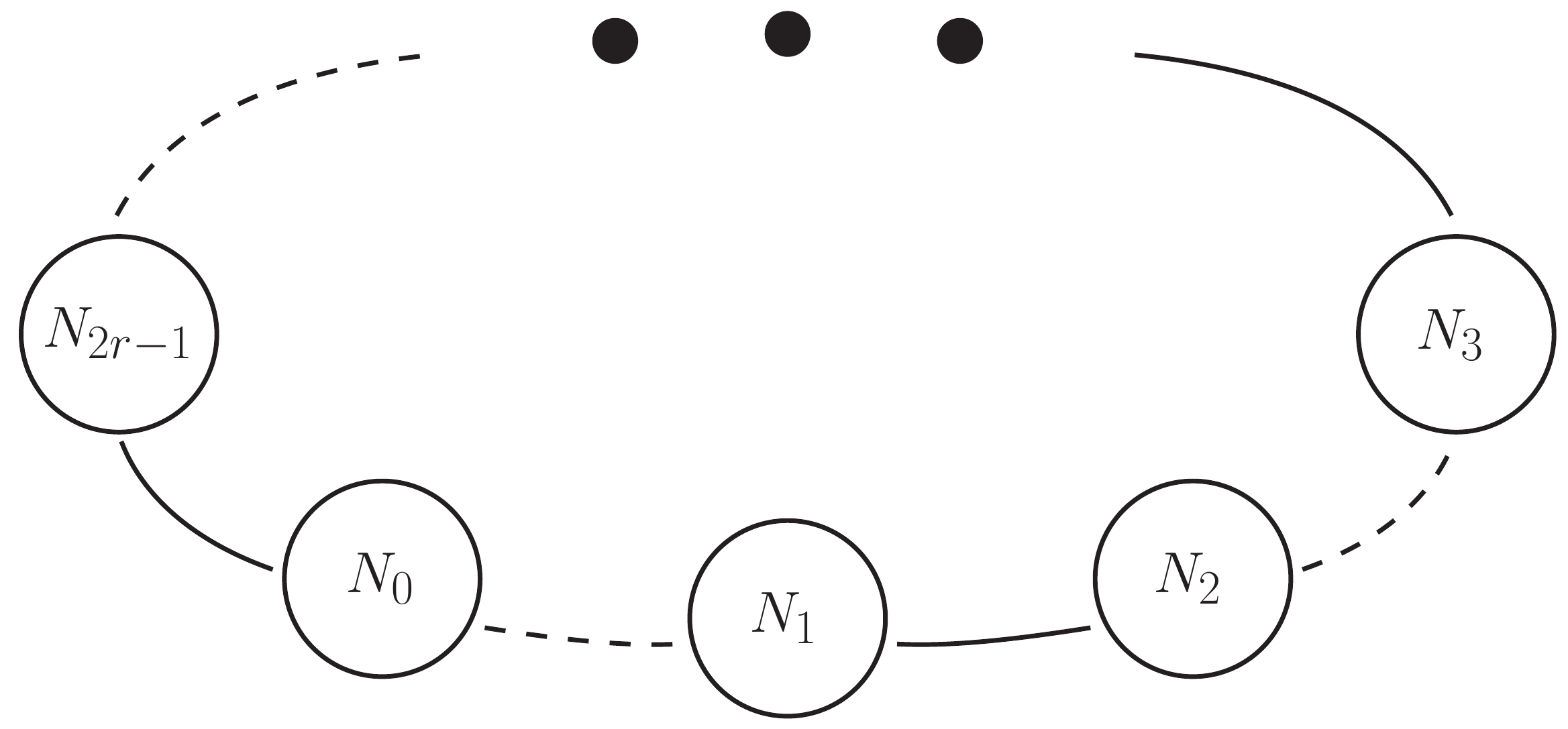}
\caption{Quiver diagram corresponding to ${\cal N}=4$ supersymmetric CS--matter theory. Solid lines represent matter hypermultiplets, while dashed lines are twisted hypermultiplets.  }
    \label{quiver}
\end{figure}

We consider a Chern--Simons--matter theory associated to a necklace quiver with gauge group $U(N_0) \times U(N_1) \times \cdots U(N_{2r-1})$ ($N_{2r} \equiv N_0$) (see Fig. \ref{quiver}).  
The field content of the theory is given by $A_{(A)}^\mu$ gauge vectors in the adjoint representation of the group $U(N_A)$ plus $r$ scalars $(q_{(2A+1)}^I)^j_{\; \hat{j}}$ ($(\bar{q}_{(2A+1) I})^{\hat{j}}_{\; j}$) in the (anti)bifundamental representation of the $U(N_{2A+1})$, $U(N_{2A+2})$ nodes (indices $j$ and $\hat j$, respectively) and in the fundamental of the R-symmetry $SU(2)_L$ ($I=1, 2$), $r$ twisted scalars $(q_{(2A)}^{\hat I})^{\hat j}_{\; j}$ ($(\bar{q}_{(2A) \hat I})^{j}_{\; \hat{j}}$) in the (anti)bifundamental representation of $U(N_{2A})$, $U(N_{2A+1})$ nodes and in the  fundamental of  
the R-symmetry $SU(2)_R$ ($\hat I=1, 2$), plus the corresponding fermions $(\psi_{(2A+1) \hat I})^j_{\; \hat{j}}$ ($(\bar{\psi}_{(2A+1)}^{\hat I})^{\hat{j}}_{\; j}$) and $(\psi_{(2A) I})^{\hat{j}}_{\; j}$ ($(\bar{\psi}_{(2A)}^I)^{j}_{\; \hat{j}}$), respectively.

The theory is ${\cal N}=4$ supersymmetric if  the CS levels satisfy the condition
\begin{equation}
k_A=\frac{k}{2}(s_A-s_{A-1}),\qquad
s_A=\pm1,\qquad
k>0
\label{ki}
\end{equation}
We will consider the case $s_A=(-1)^{A+1}$, which leads to alternating $\mp k$ levels.
Details concerning the action, the propagators and the relevant interaction vertices are given in Appendix \ref{sec:conventions}. 

This theory has a string dual description in terms of M--theory in the orbifold background AdS$_4 \times S^7/(Z_r \oplus Z_r)/Z_k$.  When $N_0 = \dots = N_{2r}$ the dual description is given by M--theory on the AdS$_4 \times S^7/(Z_r \oplus Z_{rk})$.

In analogy with the more famous examples of ABJ(M) models, bosonic BPS WL can be introduced that contain only couplings to scalars, and fermionic BPS WL that contain couplings to fermions as well. The building blocks of these operators are defined ``locally'' for each quiver node $A$ and contain matter fields 
that are at most linked to nodes $A-1$ and $A+1$. In order to simplify equations that would be otherwise cumbersome, without loosing generality we will restrict to the specific case $A=1$. 

\subsection{The bosonic 1/4 BPS WL}

Following \cite{OWZ1, Cooke:2015ila} we introduce the bosonic WL defined as 
\beq
\label{onequarter}
W_{1/4}^{+}[\G] = \frac{1}{N_1+N_2} \, {\Tr} \,P \exp{ \left( -i \int_\G d\tau {\cal L}_{1/4}^+(\tau)\right) }  \quad , \quad
{\cal L}_{1/4}^+(\tau) = \left( \begin{array}{cc}  {\cal L}_{1/4}^{(1)} & 0  \\ 0 & {\cal L}_{1/4}^{(2)} \end{array} \right) 
\eeq
where
\bea
\label{bosonicA}
&& 
{\cal L}_{1/4}^{(1)} = \dot x^\mu A_{(1) \mu} - \frac{i}{k} \left(\bar q_{(0) \hat I} (\sigma_3)^{\hat I}_{\; \hat J}  \, q_{(0)}^{\hat J}   +
q_{(1)}^I (\sigma_3)_{ I}^{\; J}  \, \bar q_{(1) \, J}    \right)  |\dot x|  
\non \\
&& 
{\cal L}_{1/4}^{(2)}= \dot x^\mu A_{(2) \mu} - \frac{i}{k} \left(\bar q_{(1) I} (\sigma_3)^{I}_{\;  J}  \, q_{(1)}^{J}  +
q_{(2)}^{\hat I} (\sigma_3)_{\hat I}^{\; \hat J} \, \bar q_{(2) \, \hat J}   \right)   |\dot x| 
\eea
Note that matter couplings involve scalars $q_{(1)}$ from the hypermultiplet connecting nodes 1 and 2 (solid line in Fig. \ref{quiver}), and scalars $q_{(0)}, q_{(2)}$ from the adjacent twisted hypermultiplets (dashed lines in Fig. \ref{quiver}).
 
The operator can be conveniently expressed in terms of WL associated to nodes 1 and 2 as
\beq
\label{Wplus}
W_{1/4}^{+} = \frac{N_1 W_{1/4}^{(1)} + N_2 W_{1/4}^{(2)}}{N_1+N_2}
\eeq
where we have defined 
\beq
\label{W12}
W_{1/4}^{(A)}[\G] = \frac{1}{N_A} \, {\Tr} \,P \exp{ \left( -i \int_\G d\tau {\cal L}^{(A)}_{1/4}(\tau)\right) }  \qquad A=1,2
\eeq
When $\G$ is a maximal circle in $S^2$ operator (\ref{onequarter}) preserves 1/4 of the supersymmetry charges. We will work in this case, parametrizing the path as  
\beq
\label{circle}
\G: \quad x^\mu (\tau) = (\cos{\tau}, \sin{\tau},0)  \qquad \; 0 \leq \tau < 2\pi
\eeq

\vskip 15pt

\subsection{The fermionic 1/2 BPS WL}

The addition of fermions leads to two inequivalent WL depending on which $SU(2)$ component we consider \cite{Cooke:2015ila}. 

The first operator, called the $\psi_1$--loop in \cite{Cooke:2015ila}, is defined in terms of $\psi_{(1) \hat 1}$ and $\bar \psi^{\hat 1}_{(1)}$ fermionic components. It is given as the generalized holonomy
\beq
W_{\psi_1}[\G] = \frac{1}{N_1 + N_2} \, {\Tr} \,P \exp{ \left( -i \int_\G d\tau {\cal L}_{\psi_1}(\tau)\right) } 
\eeq
where
\bea \label{WL1}
&& {\cal L}_{\psi_1} = \left( \begin{array}{cc} {\cal A}_{(1)}  & \bar c_\a \psi_{(1) \hat 1}^\a \\ c^\a \bar \psi^{\hat 1}_{(1) \a}  &  {\cal A}_{(2)} \end{array} \right) 
\non \\
&& {\cal A}_{(1)} =  \dot x^{\mu} A_{(1) \mu}   - \frac{i}{k} \left( q_{(1)}^I \delta_I^{\; J}  \bar q_{(1)  J} + \bar q_{(0) \hat I}  (\sigma_3)^{\hat I}_{\; \hat J}  \, q_{(0)}^{\hat J} \right) |\dot x| 
\non \\
&& {\cal A}_{(2)} =  \dot x^{\mu}  A_{(2) \mu}   - \frac{i}{k} \left(  \bar q_{(1) I}  \d^I_{\; J} \, q_{(1)}^{ J}+  q_{(2)}^{\hat I} (\sigma_3)_{\hat I}^{\; \, \hat J} \bar q_{(2) \, \hat J}  \right) |\dot x| 
\eea
and the commuting spinors $c, \bar c$ are defined in (\ref{c1}).  

We will consider the case of $\G$ being the maximal circle (\ref{circle}) for which the operator is 1/2 BPS. 
 
\vskip 10pt
An independent WL operator can be introduced that contains the $\psi_{(1) \hat 2}$ and $\bar \psi^{\hat 2}_{(1)}$ fermionic $SU(2)$ components \cite{Cooke:2015ila}. BPS invariance requires to slightly modify also the bosonic couplings, so that the $\psi_2$--loop is given by
\beq
W_{\psi_2}[\G] = \frac{1}{N_1 + N_2}  \, \Tr \,P \exp{ \left( -i \int_\G d\tau {\cal L}_{\psi_2}(\tau)\right) } 
\eeq
where
\bea\label{WL2}
&& {\cal L}_{\psi_2} = \left( \begin{array}{cc} {\cal B}_{(1)}  & \bar d_\a \psi_{(1) \hat 2}^\a \\ d^\a \bar \psi^{\hat 2}_{(1) \a}  & {\cal B}_{(2)} \end{array} \right) 
\non \\
&& {\cal B}_{(1)} =    \dot x^{\mu} A_{(1) \mu}   - \frac{i}{k} \left( -q_{(1)}^I \delta_I^{\; J}  \bar q_{(1)  J} + \bar q_{(0) \hat I}  (\sigma_3)^{\hat I}_{\; \hat J}  \, q_{(0)}^{\hat J} \right) |\dot{x}|
\non \\
&& {\cal B}_{(2)} = \dot x^{\mu}  A_{(2) \mu}   - \frac{i}{k} \left( - \bar q_{(1) I}  \d^I_{\; J} \, q_{(1)}^{ J}+  q_{(2)}^{\hat I} (\sigma_3)_{\hat I}^{\; \, \hat J} \bar q_{(2) \, \hat J}  \right)  |\dot{x}|
\eea
with the commuting spinors $d, \bar d$ given in (\ref{c2}). 

Precisely, in addition to the replacement $\psi_{(1)}^{\hat 1} \to \psi_{(1)}^{\hat 2}$ this loop differs from the previous one for $\delta_I^{\; J} \to - \delta_I^{\; J} $ in the scalar couplings and for different fermion couplings  (eq. (\ref{c1}) vs. (\ref{c2})). 
Again, when $\G$ is a maximal circle this operator is 1/2 BPS. 

\vskip 15pt

\subsection{Cohomological equivalence}

As proved in \cite{OWZ1, Cooke:2015ila}, the classical fermionic 1/2 BPS loops are both cohomologically equivalent to the 1/4 BPS bosonic operator given in eq. (\ref{Wplus}). In fact, the following relations hold
\beq
\label{coho}
W_{\psi_i} = W_{1/4}^+ + Q V_{\psi_i}  \qquad i=1,2
\eeq
where the $Q$-terms are both proportional to the same supercharge. Therefore, more generally any linear combination of the form 
\beq
\label{linear}
\frac{a_1 W_{\psi_1} + a_2 W_{\psi_2}}{a_1 + a_2}
\eeq
gives a 1/2 BPS WL that is cohomologically equivalent to the bosonic one. 

If the classical equivalence survives at quantum level, one can use $Q$ as the supercharge to localize the path integral that computes $\langle W_{1/4}^+ \rangle$ on $S^3$. As a consequence, the corresponding matrix model provides an all--order prediction not only for the bosonic $W_{1/4}^+$ but also for fermionic operators of the form (\ref{linear}), provided that they survive quantization as BPS operators. 

From the string dual description we know that at quantum level only one 1/2 BPS WL should survive, being the corresponding 1/2 BPS M2--brane configuration unique. Therefore, 
we expect that the degeneracy (\ref{linear}) gets uplifted by quantum effects and only one particular combination with fixed $\bar{a}_1, \bar{a}_2$ will correspond to the exact quantum 1/2 BPS operator. For this operator we will have
\beq
\label{cohoQ}
\langle W_{1/2} \rangle_{f=1} = \langle \frac{\bar{a}_1 W_{\psi_1} + \bar{a}_2 W_{\psi_2}}{\bar{a}_1 + \bar{a}_2} \rangle_{f=1} = \langle W_{1/4}^+ \rangle_{f=1}
\eeq
where the subscript $``f=1''$ indicates that this is the matrix model result, therefore at framing one \footnote{As discussed in \cite{Kapustin}, the Matrix Model result always refers to framing one, as the  only point--splitting regularization compatible with the supersymmetry used to localize is the one where both the original and the deformed WL contours belong to the Hopf fibration of $S^3$.}.

The uplift mechanism that breaks degeneracy at quantum level is expected to be generated by field interactions that do not occur at classical level. However, since localization actually provides the quantum exact result for the bosonic 1/4 BPS operator,  this mechanism for the fermionic ones cannot be understood within this approach.

The only possibility to disclose the degeneracy breaking mechanism is to perform a perturbative calculation of the two fermionic WL and look for potential contributions that turn out to give a different result at some loop order. In fact, if at a given order in perturbation theory we find $\langle  W_{\psi_1} \rangle \neq \langle  W_{\psi_2} \rangle$, then comparison with the localization prediction (\ref{cohoQ}) will provide a non--trivial equation that uniquely fixes the relative coefficient between $W_{\psi_1}$ and $W_{\psi_2}$, so leading to the correct quantum BPS fermionic operator. 

With this motivation in mind, we will go through the perturbative evaluation of $\langle  W_{\psi_1} \rangle$ and $\langle  W_{\psi_2} \rangle$ searching for potential differences, and match it with the weak coupling expansion of the matrix model result for $\langle W_{1/4}^+ \rangle$.

\section{All--loop relation between $W_{\psi_1}$ and $W_{\psi_2}$ }\label{sec:relation}

We approach the perturbative analysis by first deriving an all--loop identity between the  $W_{\psi_1}$ and  $W_{\psi_2}$ expectation values. 
In particular, we prove that as a consequence of the planarity of the contour $\G$ in (\ref{circle}), at a given order $L$  the two  WL are related by
\beq
\label{relation}
\langle W_{\psi_2} \rangle^{(L)} = (-1)^L \, \langle W_{\psi_1} \rangle^{(L)} 
\eeq
Here $L$ counts the power of the coupling $1/k$.  

To prove this relation, as an intermediate step we introduce a third fermionic operator that is defined from $W_{\psi_1}$ by applying a $SU(2)_L \times SU(2)_R$ transformation that exchanges the R--symmetry indices $1 \leftrightarrow 2, {\hat 1} \leftrightarrow {\hat 2}$. From the $W_{\psi_1}$ defining equations (\ref{WL1}), we then obtain a new operator $\Tilde W_{\psi_2}$ given by the holonomy of the following superconnection
\bea \label{WL2tilde}
&& {\tilde {\cal L}}_{\psi_2} = \left( \begin{array}{cc} {\cal \tilde{A}}_{(1)}  & \bar c_\a \psi_{(1) \hat 2}^\a \\ c^\a \bar \psi^{\hat 2}_{(1) \a}  &  {\cal \tilde{A}}_{(2)} \end{array} \right) 
\non \\
&& {\cal \tilde{A}}_{(1)} =  \dot x^{\mu} A_{(1) \mu}   + \frac{i}{k} \left(- q_{(1)}^I \delta_I^{\; J}  \bar q_{(1)  J} + \bar q_{(0) \hat I}  (\sigma_3)^{\hat I}_{\; \hat J}  \, q_{(0)}^{\hat J} \right) |\dot x| 
\non \\
&& {\cal \tilde{A}}_{(2)} =  \dot x^{\mu}  A_{(2) \mu}   + \frac{i}{k} \left(-  \bar q_{(1) I}  \d^I_{\; J} \, q_{(1)}^{ J} + q_{(2)}^{\hat I} (\sigma_3)_{\hat I}^{\; \, \hat J} \bar q_{(2) \, \hat J}  \right) |\dot x| 
\eea
where the commuting spinors $c, \bar c$ are still given in (\ref{c1}).  

Since the action of the theory is invariant under the R--symmetry group it is a matter of fact that computing perturbatively the expectation value of 
${\Tilde W}_{\psi_2}$ we find
\beq
\langle {\Tilde W}_{\psi_2} \rangle = \langle W_{\psi_1} \rangle
\eeq
at any given order. 

The interesting observation is that $W_{\psi_2}$ differs from $\Tilde W_{\psi_2}$ simply by an overall sign change in the scalar couplings and the replacement of the spinor couplings $c \rightarrow d$. 

\begin{figure}
\centering
 \includegraphics[width=0.25\textwidth]{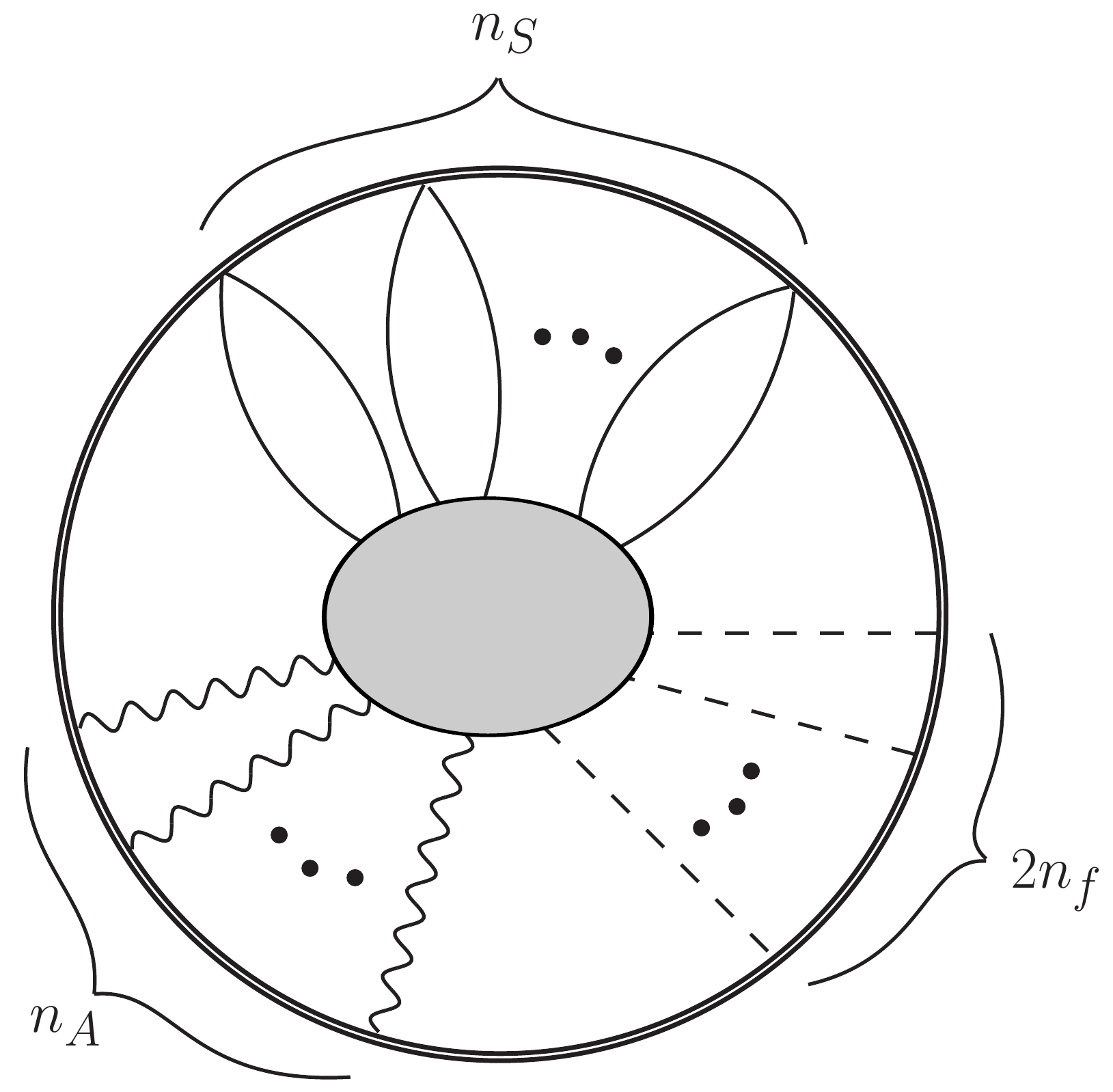}
\caption{Sketchy structure of loop diagrams contributing to the term in the WL expansion with $n_A$ gauge fields, $n_F$ $(\psi, \bar{\psi})$ couples and $n_S$ scalar bilinears. The arguments of this Section are not sensitive to the order of the contour points. }
    \label{count}
\end{figure}

Therefore, for a diagram containing $n_S$ scalar couplings from the WL expansion (see Fig. \ref{count}) the contribution to $\langle W_{\psi_2} \rangle$ is obtained from $\langle W_{\psi_1} \rangle$ simply as 
\beq
\label{ctod}
\langle W_{\psi_2} \rangle =  (-1)^{n_S} \, \langle {\Tilde W}_{\psi_2} \rangle|_{c \rightarrow d}  =  (-1)^{n_S} \, \langle W_{\psi_1} \rangle|_{c \rightarrow d}
\eeq
We now discuss what is the effect of replacing $c$ spinors with $d$ ones.  

A diagram containing $2n_F$ fermionic couplings from the $W_{\psi_1}$ expansion (see Fig. \ref{count})
is proportional to $n_F$ bilinears of the form $(c \g^{\mu_1} \g^{\mu_2} \cdots \g^{\mu_p} {\bar c})$ where the gamma matrices come from fermionic propagators, eq. (\ref{treefermion}) and gauge-fermion vertices, eq. (\ref{gaugefermion}). The gamma indices are then contracted either with external vectors, that is $x_\mu(\tau)$ or ${\dot x}_\mu(\tau)$ integrated on the contour, or with $x$--coordinates associated to internal vertices and then subject to 3D integration. According to $p$ being even or odd, using identities (\ref{id1}) for gamma matrices, the bilinears can always be reduced to linear combinations of the following structures
\bea
\label{even}
&&   (c \g^{\mu_1} \g^{\mu_2} \cdots \g^{\mu_{2m}} {\bar c}) \; \; \; \longrightarrow  \; \; (c {\bar c}) \;  \, \quad {\rm and}  \quad  \varepsilon_{\mu_i \mu_j \nu} (c \g^\n {\bar c})
\\
\label{odd}
&&  (c \g^{\mu_1} \g^{\mu_2} \cdots \g^{\mu_{2m+1}} {\bar c}) \longrightarrow  (c \g^{\mu_i} {\bar c}) \quad {\rm and} \quad  \varepsilon_{\mu_i \m_j \m_k} (c {\bar c})  
\eea
times delta and epsilon structures that account for the other $\m$--indices. 

Multiplying all the bilinears associated to a given diagram once reduced in this way, we end up with a linear combination of structures that contain powers of $(c {\bar c})$ times powers of $(c \g {\bar c})$. Let's call $n_\g$ the total number of $(c \g {\bar c})$ bilinears. 

According to the identities in Appendix \ref{sec:conventions}, these bilinears may differ at most by an overall sign when we replace $c$ with $d$ spinors. Precisely,
$(c {\bar c}) = (d {\bar d})$, $(c \g^{1,2} {\bar c}) = - (d \g^{1,2} {\bar d})$ and  $(c \g^{3} {\bar c}) = (d \g^{3} {\bar d})$. Therefore, the effect of the replacement $c \to d$ in (\ref{ctod}) will be at most an overall sign, but it is important to count how many signs we get in a given diagram. 

If we perform all Feynman integrals associated to internal vertices, before solving the contour integrals we obtain a function of the bilinears and external coordinates $x_\mu(\tau)$ and/or ${\dot x}_\mu(\tau)$. Moreover, the planarity of the contour (\ref{circle}) requires having an even number of epsilon tensors that can then be traded with products of Kronecher deltas \footnote{In fact, any string of an odd number of $\varepsilon$ tensors can be always reduced to a linear combination of products of Kronecker deltas times one epsilon tensor that would be eventually contracted with external indices, so leading to a vanishing result at framing 0.}.
It follows that  the $n_\g$ $(c \g {\bar c})$ structures end up being necessarily contracted either among themselves or with external points. However, since structures of the form $(c {\bar c})$ and $(c \g^\nu {\bar c}) (c \g_\nu {\bar c})$ do not contribute with any sign, we can restrict the discussion to the set of $(c \g {\bar c})$ contracted with external points.
Once again, the planarity of the contour (\ref{circle}) implies that the final expression will contain only bilinears of the form $(c \g^{1,2} {\bar c})$ that, according to the identities in Appendix \ref{sec:conventions}, will contribute with a sign change under replacement $c \to d$.  

From this preliminary analysis we can conclude that a given diagram containing $n_S$ scalar couplings and proportional to $n_\g$ bilinears $(c \g {\bar c})$ provides contributions to the expectation values of the two fermionic WL that are related as
\beq
\label{nn}
\langle W_{\psi_2} \rangle|_{n_S, n_\g}=  (-1)^{n_S + n_\g} \, \langle W_{\psi_1} \rangle|_{n_S, n_\g} 
\eeq
Now, combining power counting arguments with constraints coming from planarity it can be proven that  $(n_S + n_\g)$ has the same parity of the loop order $L$, or equivalently that $n_\g$ has the same parity of $L+n_S$. We leave the details of the proof of this statement in  Appendix  \ref{sec:alldetails}. 
Using this result in (\ref{nn}) we finally obtain the initial claim (\ref{relation}). 

Using similar arguments, in Appendix \ref{sec:alldetails} we also prove that all results derived perturbatively at trivial framing are real.

The loop identity \eqref{nn} implies that the expectation values of the two fermionic WL are exactly the same at any even order $L$, while they are opposite in sign at odd orders. Therefore, if quantum uplift occurs it has to be necessarily searched at odd orders. In Section \ref{lifting} we perform a systematic investigation up to $L=3$ and provide an explicit computation showing that this is the first odd order where non--vanishing (then non--trivially opposite in sign) contributions arise.

\section{The matrix model result for 1/4 BPS Wilson loop}\label{sec:MM}

The evaluation of  both the partition function and  the 1/4 BPS Wilson loop for the  necklace quiver theories  described  in Section \ref{generalities} can be
reduced to a  putative matrix integral  through localization techniques  \cite{Kapustin}.   An integral representation for the former  can be easily obtained by combining  the basic building blocks given in \cite{Kapustin}. We easily find \cite{Marino:2012az}
\be
\label{matrixmodel}
\mathcal{Z}\!=\! \mathcal{N}\!\!\int\!\! \prod_{B, i} {d\lambda_{B i} } e^{2 i k \ell_B \lambda_{Bi}^2}\!
\prod_{B=0}^{2r -1} \frac{\prod_{i<j} \sinh^2\left(\lambda_{Bi}-\lambda_{Bj}\right)}{\prod_{i,j}\cosh\left(\lambda_{Bi}-
\lambda_{B+1,j}\right)},
\ee
where we recognize the contribution of the classical action, $\prod_{Bi}e^{2 i k \ell_B \lambda_{Bi}^2}$, the one-loop fluctuations of the vector multiplets  $\prod_{i<j} \sinh^2\left(\lambda_{Bi}-\lambda_{Bj}\right)$ and those of 
the hypermultiplets $\prod_{i,j}\cosh\left(\lambda_{Bi}-
\lambda_{B+1,j}\right)$. The constant $\mathcal{N}$ is an overall normalization, whose explicit form is irrelevant in our analysis. To be consistent with the perturbative calculation we set $l_B = (-1)^B$.

In this context  the $1/4$ BPS Wilson loop is given by the vacuum expectation value of  the following matrix observable
\begin{align}
\label{Wb}
\!\!
W^{(A)}\!=&\frac{1}{N_A}\sum_{i=1}^{N_A} e^{ 2  \lambda_{Ai}}=1\!+\!\frac{2}{N_A} \text{Tr}(\Lambda_A)\!+\!\frac{2}{N_A } \text{Tr}(\Lambda^2_A)+\!\nonumber\\ &+\frac{4}{3 N_A } \text{Tr}(\Lambda_A^3)+\frac{2}{3N_A} \text{Tr}(\Lambda^4_A)+
{\mathcal O}\left(\Lambda_A^5 \right)
\end{align}
where we have introduced the diagonal matrix  $\Lambda_A\equiv\text{diag}(\lambda_{A1},\cdots,\lambda_{A N_A})$ for future convenience.  In  the r.h.s. of \eqref{Wb} we can actually neglect  all the odd powers in $\Lambda_A$  since their expectation value vanishes at all orders in $\frac{1}{k}$  due to  the symmetry property of the integrand  in  \eqref{matrixmodel} under  the parity transformation  $\lambda_{Ai}\to -\lambda_{Ai}$.

The first step to construct the perturbative series of $W^{(A)}$  is to  rescale all the eigenvalues $\lambda_{Ai}$  by  $\frac{1}{\sqrt{k}}$ and expand the integrand in \eqref{matrixmodel} for large $k$. The measure factor for large $k$ reads
\begin{align}
\label{measure}
&\prod_{A=0}^{2r -1} \frac{\prod_{i<j} \sinh^2\frac{\lambda_{Ai}-\lambda_{Aj}}{\sqrt{k}}}{\prod_{i,j}\cosh\frac{\lambda_{Ai}-
\lambda_{A+1,j}}{\sqrt{k}}}=\nonumber\\
&=\left[1+\frac{1}{k} \sum_{A=0}^{2r-1} P_A+\frac{1}{k^2} \sum_{A=0}^{2r-1} Q_A
+\frac{1}{k^3}\sum_{A=1}^{2r-1} S_A+{\mathcal O}\left(\frac{1}{k^4}\right) \right]\prod_{A=0}^{2r -1} \prod_{i<j} \frac{(\lambda_{Ai}-\lambda_{Aj})^2}{k},
\end{align}
Since we shall write the final result as a combination
of vacuum expectation values in the Gaussian matrix
model, we have chosen to use the usual Vandermonde
determinant as the reference measure.  

Order $1/k$ in the expansion \eqref{measure}  is governed by the combination
\begin{align}
P_A\equiv &\underbrace{\frac{1}{3}( N_A \mathrm{Tr}(\Lambda_A^2)-\mathrm{Tr}(\Lambda_A)^2)}_{B_2(\Lambda_A)}-
\underbrace{\frac{1}{2 }( N_{A+1}\! \mathrm{Tr}(\Lambda_A^2)\!+\!N_{A}\! \mathrm{Tr}(\Lambda_{A+1}^2)\!-\!2\mathrm{Tr}(\Lambda_A)\mathrm{Tr}(\Lambda_{A+1}))}_{C_2(\Lambda_A,\Lambda_{A+1})}.
 \end{align}
The next  order is instead controlled by $Q_A$, whose expression can be naturally written  as the sum of four different terms
 \begin{align}
 \label{QA}
 Q_A=B_4(\Lambda_A)-C_4(\Lambda_A,\Lambda_{A+1})+
 \frac{1}{2} P_A \sum_{B=0}^{2 r-1} P_B-\frac{1}{2}  [ B_2^2(\Lambda_A)-C_2^2(\Lambda_A,\Lambda_{A+1})].
 \end{align}
In  \eqref{QA} $B_4(\Lambda_A)$  is a shorthand notation for  the coefficient of $1/k^2$  when we expand  the factor in the measure due to the vector multiplet living in the node $A$. Instead   $C_4(\Lambda_A,\Lambda_{A+1})$ arises  when we expand the contribution to the measure of the  hypermultiplet  connecting the node $A$ with the node  $A+1$ at the same order. Their explicit expressions are quite cumbersome, so we report them in Appendix \ref{appendixF}. The last two terms, containing $P_A$ and $(B_2,C_2)$ respectively, originate
from lower order terms when we take the product over different nodes.
 
Finally the   explicit form $\tfrac{1}{k^3}$ term $S_A$ in \eqref{measure} is  irrelevant since it  does not affect the evaluation of the Wilson loop. In fact, its contribution cancels out with the normalization provided by the partition function.

With the help of the expansions \eqref{Wb} and  \eqref{measure}, it is straightforward to write down the expectation value of the  Wilson loop $W^{(B)}_{1/4}$ in terms of $P_A$ and $\Lambda_A$ up to $\tfrac{1}{k^3}$ order. We find
\begin{align}
\label{WWW}
\langle W^{(B)}_{1/4}\rangle=&
1\!+\!\frac{2}{N_B k} \langle \mathrm{Tr}(\Lambda^2_B)\rangle_0
\!+\! \frac{1}{N_B k^2}\Biggl[  \frac{2}{3} \langle \mathrm{Tr}(\Lambda^4_B)\rangle_0+2\sum_{A=0}^{2r-1}\left[\langle  \mathrm{Tr}(\Lambda^2_B) P_A\rangle_0-\right.\nonumber\\ 
&\left.-\langle  \mathrm{Tr}(\Lambda^2_B) \rangle_0 \langle P_A\rangle_0\right]\Biggr]+\frac{1}{N_B k^3} \left[\frac{4}{45} \langle \mathrm{Tr}(\Lambda_B^6) \rangle_0 + \frac{2}{3}\sum_{A=0}^{2 r-1}
[\langle\mathrm{Tr}(\Lambda_B^4)P_A\rangle_0-\right.\nonumber\\
&\langle\mathrm{Tr}(\Lambda_B^4)\rangle_0\langle P_A\rangle_0] + 2\sum_{A=1}^{2r-1}\bigg [
\langle\mathrm{Tr}(\Lambda_B^2) Q_A\rangle_0\!-\!\langle\mathrm{Tr}(\Lambda_B^2)\rangle_0\langle Q_A \rangle_0-\nonumber\\ &\left.-\langle
\mathrm{Tr}(\Lambda_B^2) P_A\rangle_0 \sum_{C}\langle  P_C\rangle_0 \!+\!
\langle
\mathrm{Tr}(\Lambda_B^2) \rangle_0 \langle P_A\rangle_0 \sum_{C}\langle  P_C\rangle_0 \bigg]
\right]\!\!+\! {\mathcal O}\!\left(\!\frac{1}{k^4}\!\right).
 \end{align}
where the subscript $0$ in the expectation values indicates that the average is taken in the Gaussian matrix model.
The evaluation of orders $\tfrac{1}{k}$ and $\frac{1}{k^2}$ was discussed in ref. \cite{Griguolo:2015swa} and we shall not repeat the analysis here. We simply recall the final result 
\be
\langle W^{(B)}_{1/4}\rangle=
1-\frac{i \ell_B N_B}{2 k}-\frac{1}{24 k^2}(4 N_B^2-3 N_{B-1}
   N_B-3 N_{B+1} N_B-1)+{\mathcal O}\left(\frac{1}{k^3}\right),
\ee  
which coincides with the perturbative  result  for the 1/4 BPS Wilson loops dressed with a phase corresponding to framing one \cite{Griguolo:2015swa}. 
The combination \eqref{Wplus} reads at this order
\begin{equation}
\label{MM}
\!\!\langle W_{1/4}^+ \rangle_{f=1} = 1 + i \, \frac{N_1 - N_2}{2k} - \frac{1}{24k^2}\! \left ( \!4 N_1^2 + 4 N_2^2 - 7 N_1 N_2 - 1 - 3 \frac{N_0N_1^2 + N_2^2 N_3}{N_1 + N_2} \right)\! +  {\cal O}\!\left(\frac{1}{k^3} \right)
\end{equation}

\subsection{Range--three result at three loops}

The next step is to analyze the structure of  the  $\tfrac{1}{k^3}$
contribution. An exhaustive  evaluation of all the relevant contributions in \eqref{WWW} is quite tedious and cumbersome.  However, as already mentioned, in order to  investigate the uplift of the cohomological equivalence it is sufficient to focus our attention on terms proportional to a particular color structure. A convenient choice is to look at contributions which depend on three neighboring sites $(A-1,A,A+1)$ (range--three sector).  They  can arise only from the part
not depending on $Q_A$ in the last sum in \eqref{WWW}. In fact the other terms  in \eqref{WWW} vanish unless $A=B-1$ or $A=B$ and thus they depend only on two nodes.

\noindent
Actually, most of the contributions present in the last sum in \eqref{WWW} face a similar fate   and  we 
remain with the following putative three-node term
\begin{align}
\label{sum}
&\frac{1}{N_B k^3}  \sum_{A,C}\left[ \langle\mathrm{Tr}( \Lambda_B^2 )P_A P_C\rangle_0-
\langle\mathrm{Tr}(  \Lambda_B^2)\rangle_0\langle P_A P_C\rangle_0-{2}\langle
\mathrm{Tr}(\Lambda_B^2) P_A\rangle_0\langle  P_C\rangle_0 +\right.\nn\\
&+\left. {2}
\langle
\mathrm{Tr}(\Lambda_B^2) \rangle_0 \langle P_A\rangle_0\langle  P_C\rangle_0\right]=
\frac{1}{N_B k^3}  \sum_{A,C} \langle\mathrm{Tr}( \Lambda_B^2 )P_A P_C\rangle_0^{\rm conn.}=\nonumber\\
&=\frac{2}{N_B k^3}   \langle\mathrm{Tr}( \Lambda_B^2 )P_{B-1} P_B\rangle_0^{\rm conn.},
\end{align}
since the connected correlator can be different from zero only if 
either $(A,C)=(B-1,B)$ or  $(A,C)=(B,B-1)$. If we use the explicit
expressions for $P_B$ and $P_{B-1}$, we can   easily single out
the only non-vanishing term which depends on three gauge groups. We find 
 \be
 \begin{split}
 \frac{N_{B-1}  N_{B+1}}{4 N_B k^3} &2 
\langle\mathrm{Tr}( \Lambda_B^2 ) \mathrm{Tr}( \Lambda_B^2 ) \mathrm{Tr}( \Lambda_B^2 )\rangle_0^{\rm conn.}
=-\frac{i \ell_B}{16 k^3}  N_{B-1}   N_B N_{B+1}
\end{split}
 \ee
Specializing the results at sites $A=1, 2$ and inserting in the definition (\ref{Wplus}) we finally have  
\begin{equation}
\label{MM2}
\langle W_{1/4}^+ \rangle_{f=1}^{(3)}\Big|_{\text{range 3}} = \frac{i}{16 k^3} \, \frac{N_0 N_1^2 N_2 - N_1 N_2^2 N_3}{N_1 + N_2} 
\end{equation}
We note the appearance of imaginary contributions at odd orders. As we are going to discuss in the next subsection, they can be recognized as framing contributions.

\subsection{Removing framing}

In three dimensional CS theories, expectation values of supersymmetric WL when computed via localization acquire imaginary contributions that have the interpretation of  framing effects.

This concept was originally introduced in pure CS theories in order to define a topologically invariant regularization for WL \cite{Witten}.
Precisely, it consists in a point-splitting regularization procedure based on the requirement that in correlation functions of gauge connections different gauge vectors run on auxiliary contours $\Gamma_{f}$, infinitesimally displaced from the original one.
As a consequence, WL expectation values only depend on the linking number $\chi(\G, \G_f)$ between the framing path and the WL contour via an overall phase factor that exponentiates a one--loop contribution  \cite{Witten}
\beq
\label{framing}
\langle W_{CS} \rangle = e^{i \pi \l \chi(\G, \G_f)} \, \rho(\l)  
\eeq
where $\rho$ is a framing independent function of the coupling $\l = N/k$.
The result above can be reproduced by localization for circular Wilson loops in ${\cal N}=2$ supersymmetric CS \cite{Kapustin}, where in order to preserve supersymmetry the framing contours are Hopf fibers and hence have linking number one.

For CS theories coupled to matter the identification of framing contributions in WL expectation values computed with localization and their perturbative origin is less clear.
This issue has been recently analyzed in \cite{Bianchi:2016yzj} for the 1/6 BPS WL in the ABJ(M) model. There, it has been shown that starting from three loops matter interactions induce non--trivial perturbative corrections to the one--loop framing factor in (\ref{framing}), reproducing the localization prediction at third order.  

We now apply the procedure of \cite{Bianchi:2016yzj} to ${\cal N}=4$ CS--matter theory under investigation to provide a perturbative explanation of the imaginary terms in localization results \eqref{MM} and \eqref{MM2} as coming from framing.
In order to do so, we focus on the bosonic 1/4 BPS WL $W_{1/4}^+$, whose framing contributions are easier to understand perturbatively. The cohomological equivalence \eqref{coho} then guarantees that the 1/2 BPS WL has the same expression at framing one.

At one loop framing originates by a gluon exchange diagram (as in pure CS). Using the explicit expressions in Landau gauge (see eq. (\ref{treevector})) and taking into account that $A_{(1)}$ and $A_{(2)}$ propagators differ by an overall sign, we obtain 
\bea
\langle W_{1/4}^{(A)} \rangle^{(1)} &=& i \, (-1)^{A+1}\,  \frac{N_A}{k}  \, \frac{1}{4\pi} \oint_{\G} dx^\m \oint_{\G_f} dy^\n \; \varepsilon_{\m\n\rho} \frac{(x-y)^\rho}{|x-y|^3} 
\non \\
&\equiv&  i \, (-1)^{A+1}\,  \frac{N_A}{k}  \, \chi(\G, \G_f)
\eea
where the Gauss integral is indeed proportional to the linking number between the deformed contour $\G_f$ and the original WL path $\G$.
Combining these results for $A=1,2$ according to (\ref{Wplus}) and setting $\chi(\G, \G_f) =-1$ (framing 1 in our conventions) we reproduce exactly the one--loop framing contribution in the result (\ref{MM}). 

At two loops the framing dependence of the individual 1/4 BPS bosonic WL arises from the pure gauge sector and exponentiates the one loop contribution. Adding this to the framing independent pieces and combining the WL as in \eqref{Wplus} reproduces the two-loop result from localization \eqref{MM}.
\begin{figure}
\centering
 \includegraphics[width=0.15\textwidth]{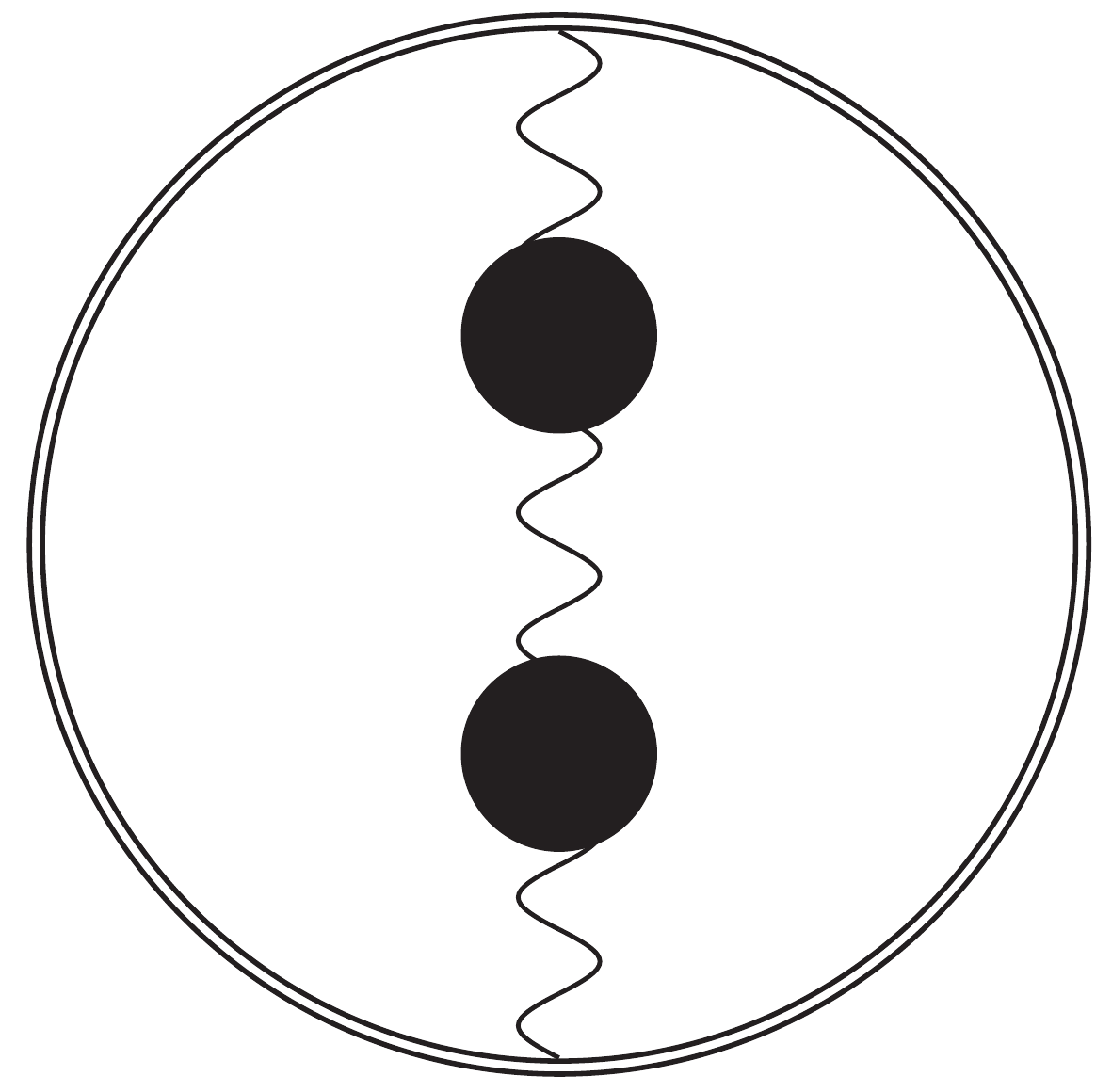}
\caption{Three--loop diagram contributing to framing. }
    \label{2loopb}
\end{figure}

At three loops, focusing on contributions in the range--three color sector, the only non--vanishing diagram is the one in Fig. \ref{2loopb}. It is associated to the exchange of one effective gauge propagator at two loops where only the one--particle reducible (1PR) corrections 
\bea
&& \langle A_{(2A+1)\mu} (x) A_{(2A+1)\nu}(y) \rangle^{(2)}_{\text 1PR} = 
- \frac{i}{4\pi}\, \frac{(N_{2A+2}+N_{2A})^2}{16\, k^3}\, \varepsilon_{\mu\nu\rho}\, \frac{(x-y)^{\rho}}{|x-y|^3} 
\non \\
&& \langle A_{(2A)\mu} (x) A_{(2A)\nu}(y) \rangle^{(2)}_{\text 1PR} = 
 \frac{i}{4\pi}\, \frac{(N_{2A+1}+N_{2A-1})^2}{16\, k^3}\, \varepsilon_{\mu\nu\rho}\, \frac{(x-y)^{\rho}}{|x-y|^3} 
\eea
can contribute with the right color structure for $A=0,1$, respectively.
The mechanism is then the same as in the one-loop computation and we obtain
\beq
\langle W_{1/4}^{(1)} \rangle^{(3)}_{\text{range 3}} =  - \frac{i}{4\pi}\, \frac{N_0 N_1 N_2}{16\, k^3}  \, \chi(\G, \G_f )
\quad \quad 
\langle W_{1/4}^{(2)} \rangle^{(3)}_{\text{range 3}} =  \frac{i}{4\pi}\, \frac{N_1 N_2 N_3}{16\, k^3}  \, \chi(\G, \G_f )
\eeq 
Combining them in $\langle W_{1/4}^+ \rangle$ and setting $\chi(\G, \G_f ) = -1$ we reproduce exactly the third order contribution (\ref{MM2}). 
We have then proved that in the matrix model result also the imaginary term (\ref{MM2}) at three loops has a framing origin. 

More generally, from the expansion of the matrix model \eqref{matrixmodel} one can argue that the expectation value of the WL is purely imaginary at odd loop orders.
On the other hand, we show in Appendix \ref{sec:alldetails} that the perturbative computation performed at trivial framing produces real terms only.
Comparing the two results we infer that all the imaginary odd order terms of the localization expression originate from framing.

The framing factor pointed out above constitutes a new kind of contribution that arises from the matter sector, in contradistinction with the pure CS phase.
We stress that such an occurrence shares the same ilk of that recently uncovered at three loops for the 1/6 BPS WL in the ABJM model in \cite{Bianchi:2016yzj} and mentioned at the beginning of this Section. In that situation an analogous 1PR diagram contributes, along with other diagrams, to reproduce the three loop imaginary term of the localization weak coupling expansion.
For the quiver theories under investigation in this paper, the possibility of distinguishing different color factors allows to single out a unique contribution from this diagram in the range-three sector, thus providing an even sharper signature of matter triggered framing phenomena.

We now turn to the fermionic 1/2 BPS operator, whose framing factor we want to isolate and remove, in order to be able to perform a comparison between the localization result and the field theory computation. 
In this case the role played by framing in fermionic diagrams is less clear.
In the context of the 1/2 BPS WL in the ABJM model it is believed that fermionic diagrams contribute to framing in such a way that its total effect exponentiates into the phase $\exp{\tfrac{i}{2} (\l_1-\l_2)}$, in agreement with the localization result \cite{Drukker:2009hy,Drukker:2010nc,Bianchi:2016gpg}.
By analogy with that picture and by comparison between the two-loop results, as carried out in \cite{Griguolo:2015swa}, we expect that the contribution of framing still exponentiates in the 1/2 BPS operator for ${\cal N}=4$ CS-matter theories. Therefore we remove the framing dependence from the localization result by taking its modulus
\beq\label{noframing}
\langle W_{1/2} \rangle_{f=0}  = 1 - \frac{1}{24k^2} \left ( N_1^2 + N_2^2 - N_1 N_2 - 1 - 3 \frac{N_0N_1^2 + N_2^2 N_3}{N_1 + N_2} \right) + {\cal O}(k^{-4})
\eeq
This expression can be checked against a three--loop perturbative calculation done in ordinary perturbation theory at framing zero.  In particular, it does not contain any third order, range--three term once the framing phase has been stripped off.

\section{Quantum uplift of cohomological equivalence}\label{lifting}

According to the cohomological arguments in Section \ref{generalities} that lead to identity (\ref{cohoQ}) and properly removing the framing factor, localization result (\ref{noframing}) should provide the expectation value at weak coupling for the actual quantum 1/2 BPS fermionic WL. In particular, this implies that while at two loops the BPS combination $\frac{(\bar{a}_1 W_{\psi_1} +\bar{a}_2 W_{\psi_2})}{\bar{a}_1 + \bar{a}_2}$ receives a non--trivial contribution, at one and three loops in the range--three color sector it should not receive any non--vanishing contribution as long as the calculation is performed at framing zero. 

On the other hand, from a perturbative perspective the general identity (\ref{relation}) tells us that computing separately $W_{\psi_1}$ and $W_{\psi_2}$, at two loops they turn out to be identical while at one and three loops non-vanishing contributions differ by an overall sign. Therefore, while no information about the actual BPS combination can be extracted at two loops, if there are non--vanishing contributions at one or three loops, matching localization and perturbative results will fix ${a}_2  = {a}_1$ in (\ref{linear}).  

This is what we are going to discuss in this Section by performing an explicit calculation at three loops.

In \cite{Griguolo:2015swa} a preliminary analysis at two loops for $W_{\psi_1}$ and $W_{\psi_2}$ has been performed using ordinary perturbation theory at framing zero. At one loop the result is zero for both WL due to the planarity of the contour, so moving to three loops the possible uplift of the classical degeneracy. 

At two loops the result reads
\begin{equation}
 \label{2loopresult}
\langle W_{\psi_1}\rangle^{(2)} = \langle W_{\psi_2} \rangle^{(2)} = -\frac{1}{24 k^2} \left[ (N_1^2 + N_2^2 - N_1 N_2 -1) - 3 \frac{N_0 N_1^2 + N_3 N_2^2}{N_1+N_2} \right]
\end{equation}
and can be used as an explicit confirmation of the general identity (\ref{relation}), besides being a non--trivial check of the matrix model result. 

At three loops, there is evidence that some diagrams are non--vanishing so they could give rise to a different result for the two WL. In \cite{Griguolo:2015swa}, a particular triangle diagram with three scalar vertices has been computed and the result turns out to be non--vanishing and opposite in sign for the two WL, in agreement with the all--loop identity (\ref{relation}).

\begin{figure}
\centering
 \includegraphics[width=0.30\textwidth]{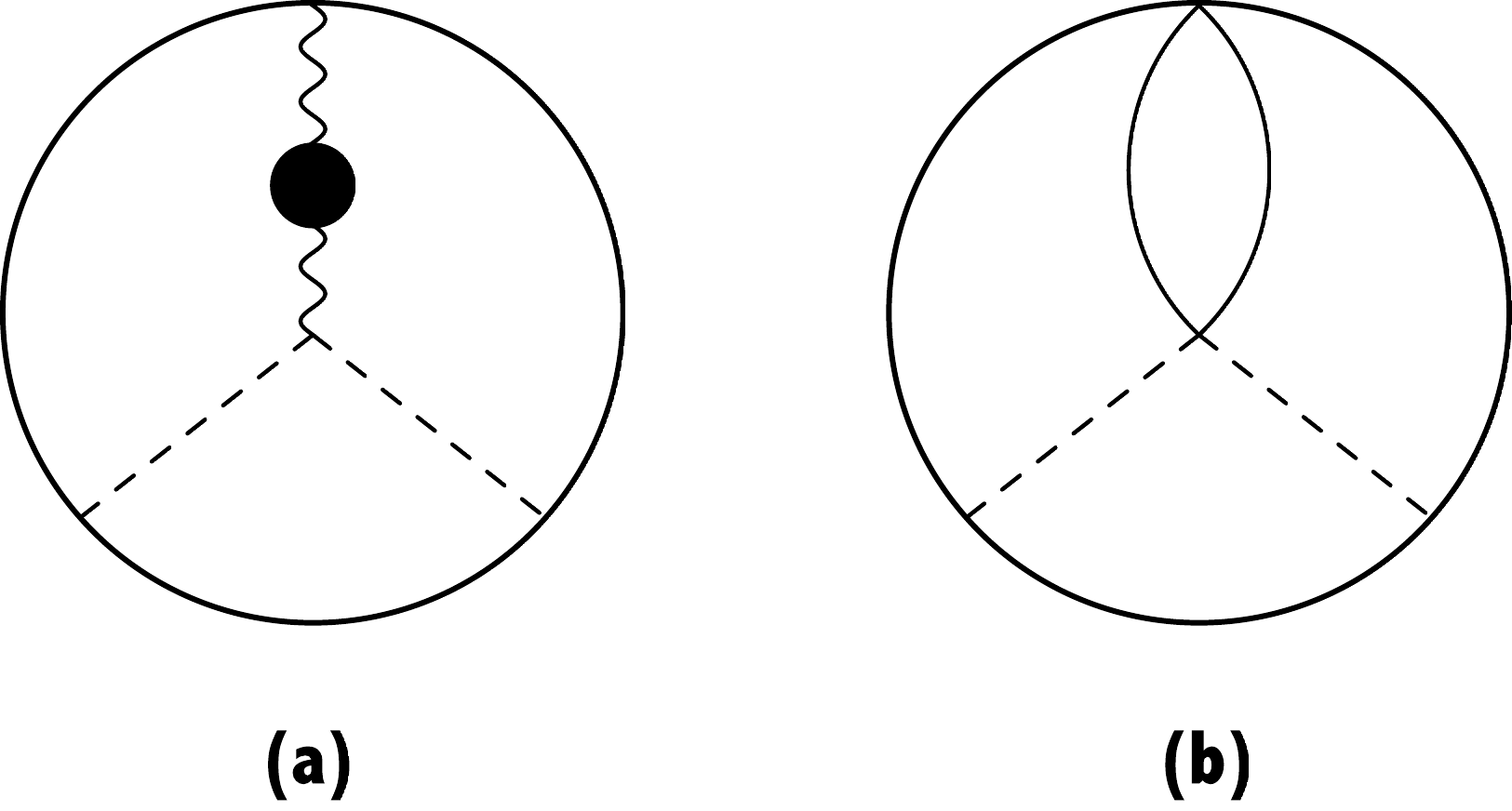}\hspace{0.7cm}\includegraphics[width=0.30\textwidth]{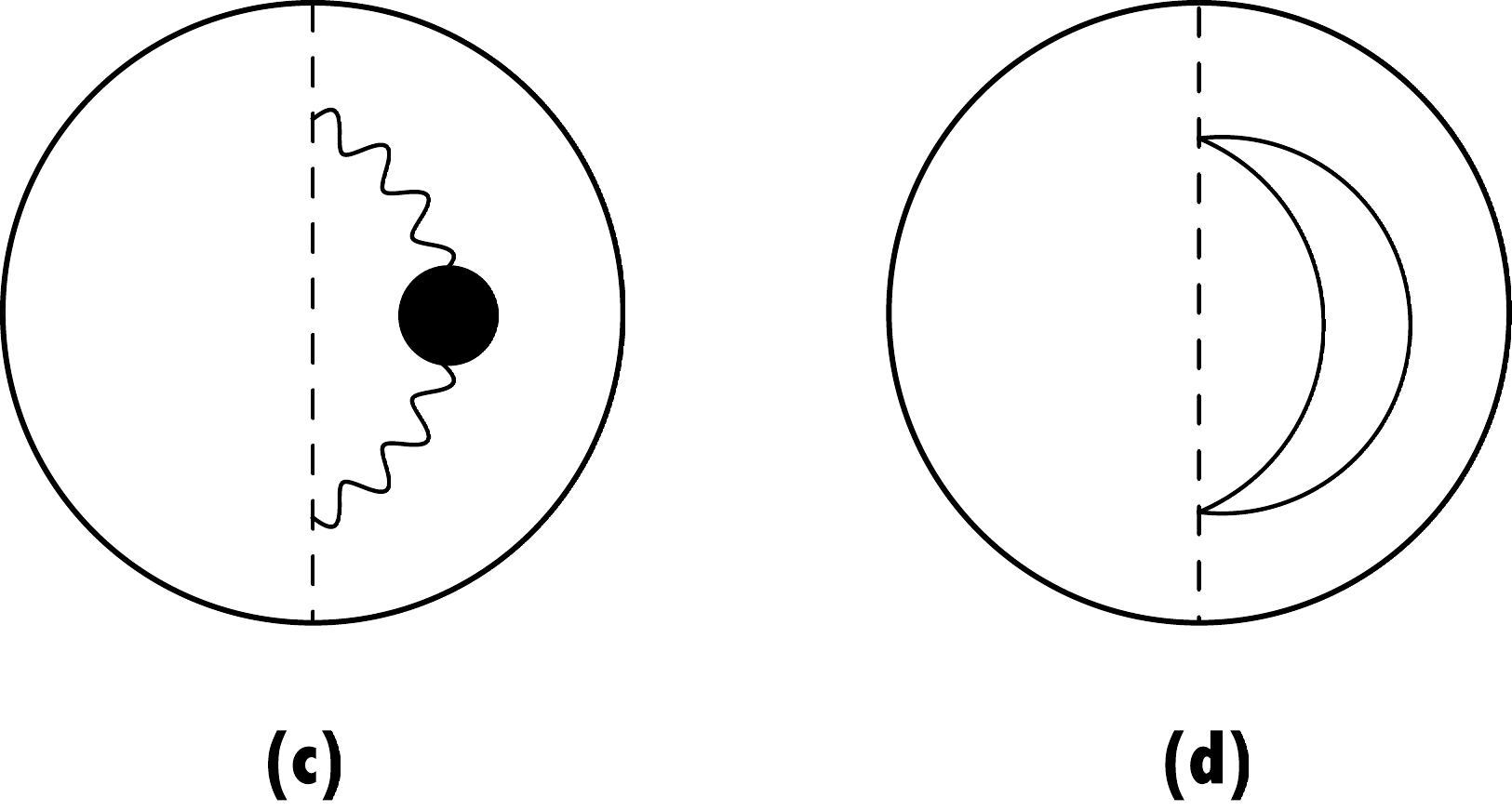}\hspace{0.7cm}\includegraphics[width=0.30\textwidth]{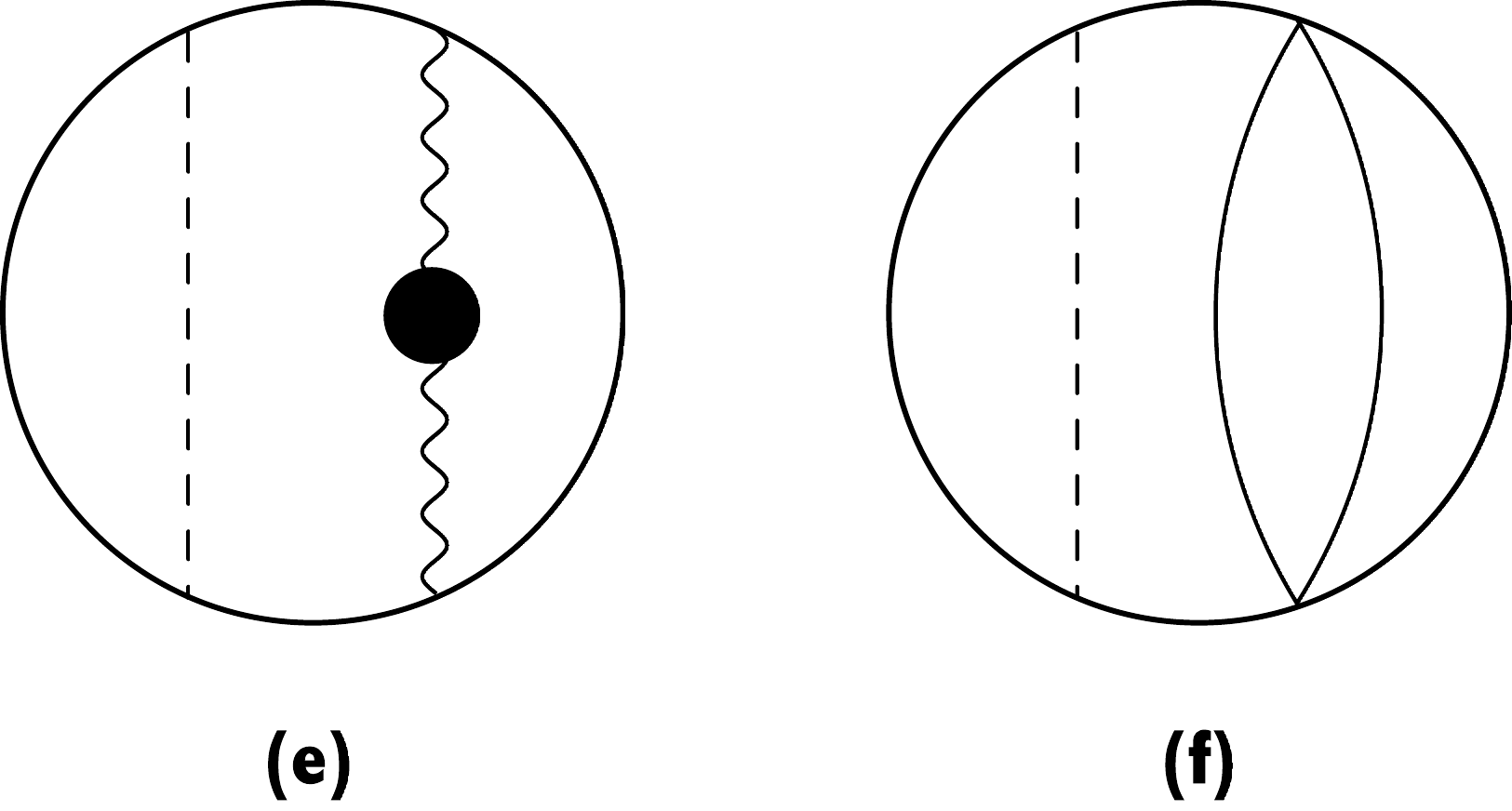}
\caption{Range--three fermionic diagrams. Black dots represent one--loop corrections to gauge propagators.}
    \label{3ldiag}
\end{figure}

Here, we perform a systematic investigation at three loops in the range--three color sector.  From a careful analysis it turns out that in this sector the only non--trivial contributions are the ones drawn in Fig. \ref{3ldiag}. Moreover, thanks to identity (\ref{relation}) we can focus only on the evaluation of $W_{\psi_1}$. 

The momentum integrals arising from diagrams in Fig. \ref{3ldiag} are in general UV divergent. We evaluate them using DRED prescription in $D=3-2\e$. This regularization has been already proved to be consistent with supersymmetry 
in three--dimensional CS theories \cite{Chen:1992ee, Bianchi:2013zda, BGLP2, GMPS, Bianchi:2014laa, Griguolo:2015swa, Bianchi:2016yzj}. 

At one loop the gauge propagator (\ref{1vector}) contains a total derivative term that could be removed by a gauge transformation. Therefore, being the WL a gauge invariant observable, we expect that this kind of contributions coming from diagrams (a), (c) and (e) sum up to zero. In the main body of the calculation we are going to neglect these terms, while we prove their actual cancellation in Appendix \ref{gaugedependent}. This is in fact a non--trivial check of the calculation. 

From the experience gained at two loops, in the calculation it is convenient to pair diagrams containing a one--loop gauge propagator with the ones where the gauge propagator is substituted by a scalar loop.  
Therefore, we are going to discuss them in pairs. 
We concentrate on contributions proportional to $N_0 N_1^2 N_2 $, since terms proportional to the other color structure $N_1 N_2^2 N_3$ can be easily inferred from the first ones. 

\paragraph{Diagrams (a) and (b).} We start by considering the first two diagrams in Fig. \ref{3ldiag} for which we need the third order expansion of the Wilson loops, which is proportional to 
\begin{align}
\label{3expansion}
  \int d\tau_{1>2>3} \,  
& {\Tr} 
\Big\{  \bar{c}_{2  } c_3  \, \langle \mathcal{A}_{(1)} (\tau_1) \psi(\tau_2) \bar{\psi} (\tau_3)   \rangle \,   
~+~ c_2\bar{c}_{3 }  \, \langle \mathcal{A}_{(2)} (\tau_1) \bar{\psi}(\tau_2)  \psi_(\tau_3) \rangle \,   
\non \\
&~~ 
+ \bar{c}_{3} c_1  \, \langle  \bar{\psi}(\tau_1) \mathcal{A}_{(1)}(\tau_2)  \psi(\tau_3) \rangle \,  
~+~ c_3 \bar{c}_{1}  \, \langle \psi(\tau_1) \mathcal{A}_{(2)} (\tau_2) \bar{\psi}(\tau_3)   \rangle \,  
\non \\
&~~
+ \bar{c}_{1 } c_2  \, \langle   \psi(\tau_1) \bar{\psi}(\tau_2)  \mathcal{A}_{(1)} (\tau_3)\rangle \, 
~+~ c_1 \bar{c}_{2}  \, \langle \bar{\psi}(\tau_1)   \psi(\tau_2) \mathcal{A}_{(2)}(\tau_3) \rangle \,  \,  \Big\} 
\end{align}
The terms involving $\mathcal{A}_{(1)}$ and $\mathcal{A}_{(2)}$  give rise to contributions to the range-three color structures $N_0 N_1^2 N_2 $ and $N_1 N_2^2 N_3$, respectively. Focusing only on the first color class, we have 
\begin{align} 
{\rm (a)}_{\psi_1} & =  C_{\rm ab} \!\int\! d\tau_{1>2>3} \bigg[ (c_3 \g_{\mu} \g_{\nu}\g_{\rho} \bar{c}_2) \dot{x}_1^{\nu}  \partial_2^{\rho} \partial_3^{\mu} \, \textrm{I(2,1,1)} - \big(1\!\rightarrow\!2\!\rightarrow\!3\!\rightarrow\!1\big) +\big(3\!\rightarrow\!2\!\rightarrow\!1\!\rightarrow\!3\big) \bigg] \label{apar} \\
{\rm (b)}_{\psi_1} & =  - C_{\rm ab} \int \!d\tau_{1>2>3}  \bigg[ (c_3 \g_{\mu}\g_{\rho} \bar{c}_2) \partial_2^{\rho} \partial_3^{\mu} \, \textrm{I(2,1,1)} - \big(1\!\rightarrow\!2\!\rightarrow\!3\!\rightarrow\!1\big) +\big(3\!\rightarrow\!2\!\rightarrow\!1\!\rightarrow\!3\big) \bigg] \label{bpar}
\end{align}
where we have defined \footnote{Along the calculation we use the shortening notations $x_{iw}^2 \equiv (x(\tau_i) - w)^2$ and $\t_{ij} \equiv (\t_i - \t_j)$.}
\beq
\label{integral}
 \textrm{I(2,1,1)}  = \int \! d^{3-2\e} w \frac{1}{(x_{1w}^2)^{1-2\e}}\frac{1}{(x_{2w}^2)^{1/2-\e}}\frac{1}{(x_{3w}^2)^{1/2-\e}}
\eeq
and 
\beq \label{CAB}
C_{\rm ab} =\frac{2 i N_0 N_1^2 N_2 }{(N_1+N_2)k^2}\left(\frac{\Gamma(\frac{1}{2}-\e)}{4 \p^{3/2-\e}}\right)^4 
\eeq
Summing the two contributions relevant simplifications occur and the remaining integrals can be computed in a completely analytical way. We refer the reader to Appendix \ref{sec:ab} for details in the resolutions of the integrals. Here we only quote the final result after expanding at small $\e$
\beq
[{\rm (a)+(b)}]_{\psi_1} = \frac{N_0 N_1^2 N_2 }{(N_1+N_2)k^3} \frac{e^{3 \g_E \e}}{4^4 \p^{1-3\e}} \bigg[\frac{16}{\e} + 16 (4+6\log 2) + {\cal O}(\e) \bigg]
\eeq
 
\vskip 10pt 
\paragraph{Diagrams (c) and (d).} These diagrams contain two--loop corrections to the fermion propagator.  In momentum space, for both flavors it is given by 
\begin{align}
&\frac{N_0 N_1}{k^2} \textrm{Tr}(\bar{\psi}_{(1)} (p)\g^\m \psi_{(1)}(-p)) \frac{p_\m}{(p^2)^{2\e}} (I_{(c)} + I_{(d)}) 
\end{align}
where 
\begin{align}
&  I_{\rm (c)} = \frac{-\csc(2\e\p)\sec(\e\p)\G(1/2-\e)}{2^{5-6\e}\p^{1/2-2\e}\G(5/2-3\e)\G(1-\e)\G(-1/2+\e)} = \frac{1}{96\p^2 \e} + \frac{3-\g_E+\log(4\p)}{48\p^2} + {\cal O}(\e) 
\end{align}
is the gauge correction expanded at small $\e$, whereas
\begin{align}
& I_{\rm (d)} = 22 \frac{1}{(4\p)^{3-2\e}} \frac{\G^3(1/2-\e)\G(2\e)}{3\G(3/2-3\e)} = 22 \bigg( \frac{1}{192\p^2 \e} + \frac{3-\g_E+\log(4\p)}{96\p^2} + {\cal O}(\e)  \bigg)
\end{align}
is the scalar correction. Here, Yukawa vertices in (\ref{yukawa}) have been used. 

We can now insert these results into the WL expression and, after integrating over the contour parameters the sum of the two integrals gives
\begin{align}
& [{\rm (c)+(d)}]_{\psi_1} =  96 \;  \frac{N_0 N_1^2 N_2 }{(N_1+N_2)k^3} \frac{e^{3 \g_E \e}}{4^4 \p^{1-3\e}} 
\end{align}
 
\vskip 10pt
\paragraph{Diagrams (e) and (f).} To compute diagram (e) and (f) we need the fourth order expansion of the WL operators that is proportional to (we consider only terms for the $N_0 N_1^2 N_2$ color structure)
\begin{align}
\label{4expansion}
\int d\tau_{1>2>3>4} \,  
{\Tr}
\Big\{ & \bar{c}_{1  } c_2  \, \langle \psi (\tau_1) \bar{\psi} (\tau_2)  \mathcal{A}_{(1)} (\tau_3) \mathcal{A}_{(1)} (\tau_4)  \rangle \,   
~+~ \bar{c}_{2  } c_3  \, \langle  \mathcal{A}_{(1)} (\tau_1)  \psi (\tau_2) \bar{\psi} (\tau_3) \mathcal{A}_{(1)} (\tau_4)  \rangle \,
\non \\ 
+ & \bar{c}_{3 } c_4  \, \langle  \mathcal{A}_{(1)} (\tau_1)\mathcal{A}_{(1)} (\tau_2)  \psi (\tau_3) \bar{\psi} (\tau_4)  \rangle \,
~+~ c_1 \bar{c}_{4}  \, \langle \bar{\psi}(\tau_1)  \mathcal{A}_{(1)}(\tau_2)  \mathcal{A}_{(1)}(\tau_3) \psi(\tau_4)  \rangle \,   \! \Big\} 
\end{align}
To evaluate diagram (e)  it is sufficient to make the substitution $\mathcal{A}_{(1)}(\tau_i) \rightarrow A_{(1) \mu}(\tau_i) \dot{x}_i^{\mu}$, whereas  for diagram (f) we take  $\mathcal{A}_{(1)}(\tau_i) \rightarrow - \frac{i}{k} (\bar{q}_{(0) \hat{I}}(\sigma_3)^{\hat{I}}_{\,\,\hat{J}} \,q_{(0)}^{\hat{J}})_{\t_i}$. Performing contractions and omitting the gauge--dependent part, for the $\psi_1$--loop  we obtain  
\bea
{\rm (e)}_{\psi_1}  = &&\,\,  C_{\rm ef}  \int d\tau_{1>2>3>4} \,\bigg[  \bigg( \sin^2 \frac{\t_{12}}{2} \bigg)^{-1+\e} \frac{\cos \t_{34}}{\big(\sin^2 \frac{\t_{34}}{2}\big)^{1-2\e}} ~+~ {\rm cyclic}  \bigg] 
\\
{\rm (f)}_{\psi_1}   =  &&\,\, -  C_{\rm ef}  \int d\tau_{1>2>3>4} \,\bigg[ \bigg( \sin^2 \frac{\t_{12}}{2} \bigg)^{-1+\e} \frac{1}{\big(\sin^2 \frac{\t_{34}}{2}\big)^{1-2\e}} ~+~ {\rm cyclic}  \bigg] 
\eea
where we have defined 
\beq \label{CEF}
C_{\rm ef} =  - \frac{N_0 N_1^2 N_2}{(N_1+N_2)k^3} \frac{\G(3/2-\e)\G^2(1/2-\e)}{2^{7-6\e}\p^{9/2-3\e}} 
\eeq
and ``+cyclic''  means $+ (1\rightarrow2\rightarrow3\rightarrow4\rightarrow1)  + (1\leftrightarrow3, 2\leftrightarrow4) +  (1\rightarrow4\rightarrow3\rightarrow2\rightarrow1)$.

Combining the two diagrams we can write
\begin{align}
[{\rm (e)+(f)}]_{\psi_1}  = &\,\,- 2 C_{\rm ef}  \int d\tau_{1>2>3>4} \,\bigg[  \bigg( \sin^2 \frac{\t_{12}}{2} \bigg)^{-1+\e}\bigg(\sin^2 \frac{\t_{34}}{2}\bigg)^{2\e} ~+~ {\rm cyclic}  \bigg]  
\non \\
 = & \frac{N_0 N_1^2 N_2 }{(N_1+N_2)k^3} \frac{e^{3 \g_E \e}}{4^4 \p^{1-3\e}} \bigg(-\frac{16}{\e} -96  \log 2 + {\cal O}(\e)  \bigg) 
\end{align}
 
 \vskip 10pt
\paragraph {The final result.} We are now ready to sum the contributions  from (a) to (f) and obtain the final result for the fermionic $\psi_1$--loop.  We note that divergent contributions from diagrams (a)+ (b) and (e) + (f) exactly cancel leading to a finite, non--vanishing result. Including also the contributions coming from the lower triangle in the WL (the ${\cal A}_{(2)}$ part), it reads
\begin{align}
\langle W_{\psi_1} \rangle^{(3)}_{\text{range 3}}    = \, \frac{5}{8 \p} \, \, \frac{N_0 N_1^2 N_2 + N_1N_2^2N_3}{(N_1+N_2)k^3}  
\end{align}
We note that this is a real result, in agreement with the general arguments of Appendix \ref{sec:alldetails} that ensure the reality of the WL expectation values at any perturbative order. Moreover, the result does not exhibit maximal transcendentality. 

According to identity (\ref{relation}) the result for the $\psi_2$--loop differs simply by an overall minus sign.  
 Therefore, if  we now consider the linear combination (\ref{linear}) at range--three we can write
\beq
\label{comb2}
\big\langle\frac{a_1 W_{\psi_1} + a_2 W_{\psi_2} }{a_1+a_2} \big\rangle^{(3)}_{\text{range 3}}  =  \frac{a_1-a_2}{a_1+a_2} \, \, \frac{5}{8 \p} \,  \frac{N_0 N_1^2 N_2 + N_1N_2^2N_3}{(N_1+N_2)k^3}  
\eeq
 The comparison with the matrix model result  cleansed from the framing contributions at three loops, eq.  \eqref{noframing}, necessarily implies $a_1 = a_2$. 
 
 We have then proved that the classical degeneracy of fermionic WL gets uplifted at three loops and the quantum 1/2 BPS WL in ${\cal N}=4$ CS--matter theories is given by
 \beq\label{eq:bps}
 W_{1/2} = \frac{W_{\psi_1} + W_{\psi_2} }{2}
\eeq

\section{Discussion}

In this paper we have identified the correct linear combination of fermionic Wilson loops that 
corresponds to the quantum 1/2 BPS operator in ${\cal N}=4$ CS--matter theories associated to necklace quivers. 
Working on the first nodes of the quiver, we have found the result in eq. (\ref{eq:bps}). The
analysis can be straightforwardly generalized to any site and we obtain
$2r$ 1/2 BPS WL with similar structure. Corresponding string solutions exist \cite{Cooke:2015ila}
and can be compared to localization predictions.

Our result solves the puzzle arisen in \cite{Cooke:2015ila}. 
The expectation value of 1/2 BPS Wilson loops in ${\cal N}=4$
CS-matter theories can be exactly evaluated through localization procedure
and reduced to a
matrix integral. The relevant configurations for
the holographic description
of 1/2 BPS Wilson loops are well understood (see \cite{Cooke:2015ila} and
reference within) and amenable, in principle, of concrete
calculations. On the field theory side the story instead is more
convoluted, due to a classical
degeneracy in the 1/2 BPS sector that seems to call for a quantum
resolution. More precisely, for circular quivers, two apparently
independent 1/2 BPS Wilson loops can be constructed at field theory level
that are indistinguishable at localization level, due to their classical
cohomological equivalence. On the other hand, at holographic level there is no
evidence of this classical degeneracy, suggesting its uplift due to
honest quantum mechanical corrections \cite{Cooke:2015ila}. Uplift is indeed detected at three loops, where the explicit 
perturbative computation distinguishes
the two different 1/2 BPS Wilson loops and only the combination (\ref{eq:bps}) coincides with the matrix integral result. 

A general analysis of the perturbative series for the two fermionic WL has revealed two important properties. 
First, there is an easy relation between the expectation values of the two operators, as they always coincide at even orders
and are opposite at odd orders. Second, the result obtained at framing zero is always real at any perturbative order. 
These properties have important consequences when we match the perturbative result with the localization prediction. In fact:

\noindent
$\bullet$ At any odd order the matrix model expansion exhibits just pure imaginary contributions. On the other hand, as we have mentioned, whatever the 1/2 BPS linear combination is, the perturbative result at framing zero is always real at any order. Matching the two results allows then to conclude that odd order terms in the localization calculation have a framing origin induced by the consistency of the procedure that necessarily require to work at framing one. We have supported this prediction with a direct three--loop calculation done at non--vanishing framing. 

Our analysis thus enlightens the role of framing in the localization procedure, extending the results of \cite{Bianchi:2016yzj} to the ${\cal N}=4$ CS-matter case. In analogy with the ABJ(M) case, we expect the framing contributions to exponentiate, so that the expectation values of WL at framing zero should be obtained by taking the modulus of the matrix model expansion. In particular, this implies that the correct quantum BPS operators have vanishing contributions at odd orders if computed in ordinary perturbation theory with no framing.

\noindent
$\bullet$
The all--loop relation between the expectation values of the two WL, eq. (\ref{relation}), suggests that potential uplifts can arise only at odd orders, if non--vanishing contributions appear there. As we have discussed in this paper, three loops is indeed the first odd order where this happens. There, the request to have a three--loop vanishing contribution to $\langle W_{1/2} \rangle$ at framing zero,  as suggested by the localization prediction, necessarily leads to the conclusion that the average (\ref{eq:bps}) is the correct combination where unwanted terms cancel.  

More generally, the arguments above allow to conclude that \eqref{eq:bps} is the exact 1/2 BPS operator at all-loop orders. In fact, whatever the non--vanishing contributions are that appear at higher odd orders for the two WL, they will be always real and opposite in sign. The linear combination  (\ref{eq:bps}) is then the only one that has vanishing odd--order terms. 
 
 \vskip 10pt
 
We have taken advantage
of working with different gauge groups in each site. This has allowed to
focus only on one specific color sector where the number of non--vanishing
diagrams is reasonably small. We cannot easily conclude anything in the
orbifold case ($N_0 = N_1 =...$) \cite{Benna:2008zy} since contributions from all the other
sectors should be included. In particular, we cannot conclude that at
three-loops we obtain a non-vanishing result, although it seems quite natural. We
remark that in this case an elegant formulation of the
theory also exists in terms of a Fermi-gas description \cite{Marino:2011eh}, which allows for efficient Wilson
loop average computations. It would be nice to identify suitable limits
that admit all-order comparisons with perturbation theory.

Our results indicate that the straightforward localization procedure
hides sometimes delicate questions regarding the quantum nature of
(composite) field operators and the choice of a regularization scheme. In
the present case, while combination $\frac{1}{2}(W_{\psi_1} + W_{\psi_2})$
is enhanced to a true 1/2 BPS operator with a well-defined
holographic dual, the other independent combination $(W_{\psi_1} -
W_{\psi_2})$ would deserve a closer inspection. This operator seems not to be 1/2 BPS
and not detectable by localization. Although it is cohomologically trivial at classical level, its expectation value is non--vanishing at three loops, it is
real and, quite unexpectedly, of lower transcendentality (see eq. (\ref{comb2})). Moreover,
it is reasonable to expect that it will be non--trivially corrected also at higher orders and
from our general power counting arguments the complete result at framing zero should be a real 
function of the couplings given by an odd-order expansion. We do not have {\em a priori} arguments to exclude the appearance of divergent contributions. 
However, our three--loop calculation seems to suggest that divergences might be absent, given that at this order the two fermionic WL turn out to be separately finite. This might be an indication that some supersymmetry survives.
It would be interesting to further investigate the physical
meaning of this operator and find its dual brane configuration.

\vfill \newpage
\appendix

\section{Conventions and Feynman rules}\label{sec:conventions}

We work in euclidean three--dimensional space with coordinates $x^\mu = (x^1, x^2, x^3)$. The set of gamma matrices satisfying $\{ \g^\mu , \g^\nu \} = 2\delta^{\mu \nu}$ is chosen to be 
\beq
(\g^\mu)_\a^{\; \, \b} = \{\s^3, \s^1,\s^2 \}
\eeq
with matrix product 
\beq
\label{prod}
(\g^\mu \g^\nu)_\a^{\; \, \b} \equiv (\g^\mu)_\a^{\; \, \g} (\g^\nu)_\g^{\; \, \b}
\eeq
Useful identities are  
\bea 
\label{id1}
&&  \g^\mu \g^\nu = \d^{\mu \nu} \mathbb{I} + i \varepsilon^{\mu\nu\rho} \g^\rho
\non \\
&& \g^\mu \g^\nu \g^\rho = \d^{\mu\nu} \g^\rho - \d^{\mu\rho} \g^\nu+  \d^{\nu\rho} \g^\mu  + i \varepsilon^{\mu\nu\rho} \mathbb{I}
\non \\
&& 
\g^\mu \g^\nu \g^\rho \g^\s -  \g^\s \g^\rho \g^\nu \g^\mu = 2i \left( \d^{\mu\nu} \varepsilon^{\rho\s \eta}  + \d^{\rho \s}  \varepsilon^{\mu\nu\eta} + \d^{\nu\eta} \varepsilon^{\rho \mu \s} +
\d^{\mu\eta} \varepsilon^{\nu\rho\s}  \right) \g^\eta
 \\
&& 
\non \\
&&
\Tr (\g^\mu \g^\nu) = 2 \d^{\mu\nu}
\non \\
&&
\Tr (\g^\mu \g^\nu \g^\rho) = 2i \varepsilon^{\mu\nu\rho}
\eea 
Spinorial indices are lowered and raised as $(\g^\mu)^\a_{\; \, \b} = \varepsilon^{\a \g}  (\g^\mu)_\g^{\; \, \d} \varepsilon_{\b \d}$, where
\beq
\varepsilon^{\a\b} =  \left( \begin{array}{cc} 0 & 1 \\ -1 & 0 \end{array} \right) 
\qquad \qquad 
\varepsilon_{\a\b} =  \left( \begin{array}{cc} 0 & -1 \\ 1 & 0 \end{array} \right) 
\eeq
It follows that 
\beq
(\g^\mu)^\a_{\; \, \b} = \{  \s^3, \s^1, -\s^2 \}  
\eeq
In addition,
\bea
&& (\g^\mu)_{\a \b} = \{  \s^1, \s^3, i \mathbb{I} \} = (\g^\mu)_{\b\a}  
\non \\
&& (\g^\mu)^{\a \b} = \{- \s^1, \s^3, i \mathbb{I} \} = (\g^\mu)^{\b\a}  
\eea
are symmetric matrices.

We conventionally choose the spinorial indices of chiral fermions to be always up, while the ones of antichirals to be always down. Therefore
\beq
({\eta}_1 \g^\mu \bar\eta_2)  \equiv ({\eta}_1^\a  (\g^\mu)_\a^{\; \, \b}  \bar\eta_{2 \;\b})  
\eeq
 
\vskip 10pt

In order to study BPS WL in ${\cal N} = 4$ supersymmetric Chern--Simons--matter theories associated to linear quivers it is sufficient to concentrate "locally" on three quiver nodes $U(N_0) \times U(N_1) \times U(N_2)$. We will then consider the gauge-matter theory for this group. 

The action relevant for two-loop calculations is ($\G = \int e^{-S}$)
\beq
S = S_{CS} + S_{matter} + S_{gf} 
\eeq
\bea
\label{action}
S_{CS} &=& - \frac{i}{2} k \, \int d^3x\,\varepsilon^{\mu\nu\rho} \Big[ {\Tr} \left( A_{(1)\mu} \partial_\nu A_{(1)\rho}+\frac{2}{3} i A_{(1)\mu} A_{(1)\nu} A_{(1)\rho} \right)
 \\
&~& \qquad \qquad \qquad \qquad \quad - {\Tr} \left({A}_{(0)\mu}\partial_\nu 
{A}_{(0)\rho}+\frac{2}{3} i {A}_{(0)\mu} {A}_{(0)\nu} {A}_{(0)\rho} \right) 
\non \\
&~& \qquad \qquad \qquad \qquad \quad - {\Tr} \left({A}_{(2)\mu}\partial_\nu 
{A}_{(2)\rho}+\frac{2}{3} i {A}_{(2)\mu} {A}_{(2)\nu} {A}_{(2)\rho} \right) 
 \Big]
\non \\
S_{matter} &=& \int d^3x \, {\Tr} \Big[  D_\mu q_{(0)}^{\hat I} D^\mu \bar{q}_{(0) \hat I}+ i \, \bar{\psi}_{(0)}^{I}  \g^\mu D_\mu \psi_{(0) I} 
\non \\
&~& \qquad \qquad + \, D_\mu q_{(1)}^{I} D^\mu \bar{q}_{(1)  I}+ i \, \bar{\psi}_{(1)}^{\hat I}  \g^\mu D_\mu \psi_{(1) \hat I} 
\non  \\
&~& \qquad \qquad +\, D_\mu q_{(2)}^{\hat I} D^\mu \bar{q}_{(2) \hat I}+ i \, \bar{\psi}_{(2)}^{I}  \g^\mu D_\mu \psi_{(2) I} 
\Big] + S_{int} 
\non \\
S_{gf} &=& \frac{k}{2} \int d^3x \, {\Tr} \Big[ \frac{1}{\xi_{(1)}}  (\pa_\mu A_{(1)}^\mu)^2 + \pa_\mu \bar{c}_{(1)} D^\mu c_{(1)}  - 
\frac{1}{\xi_{(0)}} ( \pa_\mu {A}_{(0)}^\mu )^2 - \pa_\mu \bar{\hat{c}}_{(0)} D^\mu \hat{c}_{(0)} 
\non \\
&~& \qquad \qquad \qquad - \frac{1}{\xi_{(2)}} ( \pa_\mu  {A}_{(2)}^\mu )^2 - \pa_\mu \bar{\hat{c}}_{(2)} D^\mu \hat{c}_{(2)}
\Big] \non
\eea
where $(q_{(2A+1)}^I)^j_{\; \hat{j}}$ ($(\bar{q}_{(2A+1) I})^{\hat{j}}_{\; j}$), $I=1, 2$, are matter scalars in the bifundamental (antibifundamental) representation of the $(2A+1)$, ${\hat{(2A+2)}}$ nodes and in the fundamental repr. of the R-symmetry $SU(2)_L$, whereas $(q_{(2A)}^{\hat I})^{\hat j}_{\; j}$ ($(\bar{q}_{(2A) \hat I})^{j}_{\; \hat{j}}$), $\hat I=1, 2$  are twisted scalars in the bifundamental representation of $\hat{(2A)}$, $(2A+1)$ nodes and in the  fundamental repr. of  
the R-symmetry $SU(2)_R$.  Analogously, $(\psi_{(2A+1) \hat I})^j_{\; \hat{j}}$ ($(\bar{\psi}_{(2A+1)}^{\hat I})^{\hat{j}}_{\; j}$) and $(\psi_{(2A) I})^{\hat{j}}_{\; j}$ ($(\bar{\psi}_{(2A)}^I)^{j}_{\; \hat{j}}$) describe the corresponding fermions.

The covariant derivatives are  defined as ($A=0,1$)
\bea
\label{covariant}
D_\mu q_{(2A)}^{\hat I} &=& \pa_\mu q_{(2A)}^{\hat I}  + i {A}_{(2A)\mu} q_{(2A)}^{\hat I}  - i  q_{(2A)}^{\hat I} A_{(2A+1)\mu}    
\non \\
D_\mu q_{(2A+1)}^{I} &=& \pa_\mu q_{(2A+1)}^{I}  + i A_{(2A+1)\mu} q_{(2A+1)}^{I}  - i  q_{(2A+1)}^{I} {A}_{(2A+2)\mu}  
\non \\
D_\mu \psi_{(2A) I} &=& \pa_\mu \psi_{(2A) I}  + i   {A}_{(2A) \mu} \psi_{(2A) I}  - i  \psi_{(2A) I}  A_{(2A+1) \mu}  
\non \\
D_\mu \psi_{(2A+1) \hat I} &=& \pa_\mu \psi_{(2A+1) \hat I}  + i   A_{(2A+1) \mu} \psi_{(2A+1) \hat I} - i \psi_{(2A+1) \hat I} {A}_{(2A+2) \mu} 
\eea
\bea
D_\mu \bar{q}_{(2A) \hat I} &=& \pa_\mu  \bar{q}_{(2A) \hat I}  - i \bar{q}_{(2A) \hat I}  {A}_{(2A) \mu} + i A_{(2A+1) \mu}  \bar{q}_{(2A) \hat I} 
\non \\
D_\mu \bar{q}_{(2A+1) I} &=& \pa_\mu \bar{q}_{(2A+1) I} - i  \bar{q}_{(2A+1) I} A_{(2A+1) \mu} + i {A}_{(2A+2) \mu}  \bar{q}_{(2A+1) I}
\non \\
D_\mu \bar{\psi}^I_{(2A)}  &=& \pa_\mu \bar{\psi}^I_{(2A)} - i \bar{\psi}^I_{(2A)} {A}_{(2A) \mu}  + i   A_{(2A+1) \mu} \bar{\psi}^I_{(2A)}
\non \\
D_\mu \bar{\psi}^{\hat I}_{(2A+1)}  &=& \pa_\mu \bar{\psi}^{\hat I}_{(2A+1)} - i \bar{\psi}^{\hat I}_{(2A+1)} A_{(2A+1) \mu} + i {A}_{(2A+2) \mu} \bar{\psi}^{\hat I}_{(2A+1)} 
\eea

\vskip 10pt

\noindent
From the action (\ref{action}) we obtain the following Feynman rules: 

\vskip 15pt
\noindent
\underline{The propagators} \\

\noindent Tree--level vector propagators in Landau gauge
\bea
\label{treevector}
\langle  (A_{(2A+1)\mu})^i_{\, j} (x) (A_{(2A+1)\nu})^k_{\; l}(y)  \rangle^{(0)} &=&  \d^i_{\, l} \d^k_{\; j}   \,  \frac{i}{k} \, \frac{\G(\frac32-\e)}{2\pi^{\frac32 -\e}} \varepsilon_{\mu\nu\rho} \frac{(x-y)^\rho}{[(x-y)^2]^{\frac32 -\e} }
\non \\
&=&  \d^i_{\, l} \d^k_{\; j}    \,  \frac{1}{k} \,\varepsilon_{\mu\nu\rho} \, \int \frac{d^np}{(2\pi)^n} \frac{p^\rho}{p^2} e^{ip(x-y)}
\non \\
&& \qquad 
\non \\
\langle ({A}_{(2A)\mu})^{\hat i}_{\, \hat{j}} (x) ({A}_{(2A)\nu})^{\hat k}_{\; \hat{l}} (y) \rangle^{(0)} &=&  -  \d^{\hat i}_{\, \hat{l}} \d^{\hat k}_{\; \hat{j}}    \,  \frac{i}{k} \, \frac{\G(\frac32-\e)}{2\pi^{\frac32 -\e}} \varepsilon_{\mu\nu\rho} \frac{(x-y)^\rho}{[(x-y)^2]^{\frac32 -\e} }
\non \\
&=& -
  \d^{\hat i}_{\, \hat{l}} \d^{\hat k}_{\; \hat{j}}   \,  \frac{1}{k} \,\varepsilon_{\mu\nu\rho} \, \int \frac{d^np}{(2\pi)^n} \frac{p^\rho}{p^2} e^{ip(x-y)}
\eea

\noindent One--loop vector propagators  
\bea
\label{1vector}
&& \langle   (A_{(2A+1)\mu})^i_{\, j} (x) (A_{(2A+1)\nu})^k_{\; l}(y)   \rangle^{(1)} =
\non \\
&~& \quad =  \d^i_{\, l} \d^k_{\; j}     \frac{(N_{2A} + N_{2A+2} )}{k^2} \,   \frac{\G^2(\frac12-\e)}{8\pi^{3 -2\e}} 
\,  \left[ \frac{\d_{\mu\nu}}{ [(x- y)^2]^{1-2\e}} - \pa_\mu \pa_\nu \frac{[(x-y)^2]^{2\e}}{4\e(1+2\e)} \right]  \non  \\
\non \\
&~&   \quad  =  \d^i_{\, l} \d^k_{\; j}     \frac{(N_{2A} + N_{2A+2} )}{k^2} \,   \frac{\G^2(\frac12-\e) \G(\frac12 + \e)}{ 2^{2-2\e} \pi^{\frac32 -\e}\G(1-2\e)} 
\, \int \frac{d^np}{(2\pi)^n} \frac{e^{ip(x-y)}}{(p^2)^{\frac12 + \e}} \left( \d_{\mu\nu} - \frac{p_\mu p_\nu}{p^2} \right)
\non \\
\non \\
\non \\
&& \langle ({A}_{(2A)\mu})^{\hat i}_{\, \hat{j}} (x) ({A}_{(2A)\nu})^{\hat k}_{\; \hat{l}} (y)  \rangle^{(1)} =
\non \\
&~& \quad =   \d^{\hat i}_{\, \hat{l}} \d^{\hat k}_{\; \hat{j}}    \frac{(N_{2A-1} + N_{2A+1}) }{k^2}   \frac{\G^2(\frac12-\e)}{8\pi^{3 -2\e}} 
\left[ \frac{\d_{\mu\nu}}{ [(x- y)^2]^{1-2\e}} - \pa_\mu \pa_\nu \frac{[(x-y)^2]^{2\e}}{4\e(1+2\e)} \right] \non \\
&~&   \quad =   \d^{\hat i}_{\, \hat{l}} \d^{\hat k}_{\; \hat{j}}   \frac{(N_{2A-1} + N_{2A+1} )}{k^2} \,   \frac{\G^2(\frac12-\e) \G(\frac12 + \e)}{  2^{2-2\e} \pi^{\frac32 -\e}\G(1-2\e)} 
\, \int \frac{d^np}{(2\pi)^n} \frac{e^{ip(x-y)}}{(p^2)^{\frac12 + \e}} \left( \d_{\mu\nu} - \frac{p_\mu p_\nu}{p^2} \right)
\non \\
\eea
\noindent Scalar propagator
\bea
\label{scalar}
\langle (q_{(2A)}^{\hat I})_i^{\; \hat{j}} (x) (\bar{q}_{(2A) \hat J})_{\hat{k}}^l(y) \rangle^{(0)}  &=& \d_{\hat J}^{\hat I} \d_i^l \d_{\hat{k}}^{\hat{j}} \, \frac{\G(\frac12 -\e)}{4\pi^{\frac32-\e}} 
\, \frac{1}{[(x-y)^2]^{\frac12 -\e}}
\non \\
&=& \d_{\hat J}^{\hat I} \d_i^l \d_{\hat{k}}^{\hat{j}} \, \int \frac{d^np}{(2\pi)^n} \frac{e^{ip(x-y)}}{p^2}
\\
\langle (q_{(2A+1)}^{I})_i^{\; \hat{j}} (x) (\bar{q}_{(2A+1) J})_{\hat{k}}^l(y) \rangle^{(0)}  &=& \d_{J}^{I} \d_i^l \d_{\hat{k}}^{\hat{j}} \, \frac{\G(\frac12 -\e)}{4\pi^{\frac32-\e}} 
\, \frac{1}{[(x-y)^2]^{\frac12 -\e}} 
\non \\
&=& \d_{J}^{I} \d_i^l \d_{\hat{k}}^{\hat{j}} \, \int \frac{d^np}{(2\pi)^n} \frac{e^{ip(x-y)}}{p^2}
\eea
\noindent Tree--level fermion propagator
\bea
&& \label{treefermion}
\langle (\psi_{(2A)I}^\a)_{\hat{i}}^{\; j}  (x) (\bar{\psi}_{(2A) \, \b}^J )_k^{\; \hat{l}}(y) \rangle^{(0)} = - i \, \d_I^J \d_{\hat{i}}^{\hat{l}} \d_{k}^{j} \, 
\frac{\G(\frac32 - \e)}{2\pi^{\frac32 -\e}} \,  \frac{(\g^\mu)^\a_{\; \, \b} \,  (x-y)_\mu}{[(x-y)^2]^{\frac32 - \e}}
\non \\
&& \qquad \qquad \qquad \qquad \qquad \qquad ~ =   -  \, \d_I^J \d_{\hat{i}}^{\hat{l}} \d_{k}^{j} \, (\g^\mu)^\a_{\; \, \b}   \int \frac{d^np}{(2\pi)^n} \frac{p_\mu}{p^2} e^{ip(x-y)}
\non \\
&& 
\\
&& 
\langle (\psi_{(2A+1) \hat I}^\a)_{\hat{i}}^{\; j}  (x) (\bar{\psi}^{\hat J}_{(2A+1) \, \b} )_k^{\; \hat{l}}(y) \rangle^{(0)} = - i \, \d_{\hat I}^{\hat J} \d_{\hat{i}}^{\hat{l}} \d_{k}^{j} \, 
\frac{\G(\frac32 - \e)}{2\pi^{\frac32 -\e}} \,  \frac{(\g^\mu)^\a_{\; \, \b} \,  (x-y)_\mu}{[(x-y)^2]^{\frac32 - \e}}
\non \\
&& \qquad \qquad \qquad \qquad \qquad \qquad \qquad  =   -  \, \d_{\hat I}^{\hat J} \d_{\hat{i}}^{\hat{l}} \d_{k}^{j}  \, (\g^\mu)^\a_{\; \, \b}   \int \frac{d^np}{(2\pi)^n} \frac{p_\mu}{p^2} e^{ip(x-y)}
\non \\
\eea
\noindent One--loop fermion propagator 
\bea
\label{1fermion}
&& \langle (\psi_{(2A)I}^\a)_{\hat{i}}^{\; j}  (x) (\bar{\psi}_{(2A) \, \b}^J )_k^{\; \hat{l}}(y) \rangle^{(1)} =
\non \\
&~& \qquad \qquad  =  \frac{i}{k}  \,  \d_I^J \d_{\hat{i}}^{\hat{l}} \d_{k}^{j} \,  \, \d^\a_{\; \, \b}
\, (N_{2A+1}-N_{2A}) \frac{\G^2(\frac12 - \e)}{16 \pi^{3-2\e}} \, \frac{1}{[(x-y)^2]^{1 - 2\e}}  
\non \\
&~& \qquad \qquad =  \frac{i}{k}  \,  \d_I^J \d_{\hat{i}}^{\hat{l}} \d_{k}^{j} \,  \, \d^\a_{\; \, \b} \, (N_{2A+1}-N_{2A}) \, 
\frac{\G^2(\frac12-\e) \G(\frac12 + \e)}{  2^{3-2\e} \pi^{\frac32 -\e}\G(1-2\e)} 
\, \int \frac{d^np}{(2\pi)^n} \frac{e^{ip(x-y)}}{(p^2)^{\frac12 + \e}}  
\non \\
\\
&& \langle (\psi_{(2A+1) \hat I}^\a)_{\hat{i}}^{\; j}  (x) (\bar{\psi}^{\hat J}_{(2A+1) \, \b} )_k^{\; \hat{l}}(y) \rangle^{(1)} =
\non \\
&~& \qquad \qquad   = \frac{i}{k}  \,  \d_I^J \d_{\hat{i}}^{\hat{l}} \d_{k}^{j} \,  \, \d^\a_{\; \, \b}
\, (N_{2A+1}-N_{2A+2}) \frac{\G^2(\frac12 - \e)}{16 \pi^{3-2\e}} \, \frac{1}{[(x-y)^2]^{1 - 2\e}} 
\non \\
&~& \qquad \qquad = \frac{i}{k}  \,  \d_I^J \d_{\hat{i}}^{\hat{l}} \d_{k}^{j} \,  \, \d^\a_{\; \, \b} \, (N_{2A+1}-N_{2A+2}) \, 
\frac{\G^2(\frac12-\e) \G(\frac12 + \e)}{  2^{3-2\e} \pi^{\frac32 -\e}\G(1-2\e)} 
\, \int \frac{d^np}{(2\pi)^n} \frac{e^{ip(x-y)}}{(p^2)^{\frac12 + \e}}  
\non \\
\eea	 
\vskip 15pt
\noindent
\underline{The interaction vertices}\\

\noindent 1) Gauge cubic vertices (from $(-S)$)
\bea
\label{gaugecubic}
&&- \frac{k}{3} \, \varepsilon^{\mu\nu\rho} \int d^3x \,( A_{(1)\mu})^i_{\, j} (A_{(1)\nu})^j_{\, k} (A_{(1)\rho})^k_{\; i}  
\\
&&  \frac{k}{3} \, \varepsilon^{\mu\nu\rho} \int d^3x \,  (A_{(0)\mu})^{\hat i}_{\, \hat{j}}   (A_{(0)\nu})^{\hat{j}}_{\, \hat{k}}  (A_{(0)\rho})^{\hat{k}}_{\; \hat{i}}  \quad , \quad 
 \frac{k}{3} \, \varepsilon^{\mu\nu\rho} \int d^3x \,  (A_{(2)\mu})^{\hat i}_{\, \hat{j}}   (A_{(2)\nu})^{\hat{j}}_{\, \hat{k}}  (A_{(2)\rho})^{\hat{k}}_{\; \hat{i}}   
\non
\eea
\noindent 2) Gauge--fermion cubic vertex from $(-S)$ (we only need $\psi_{(1)}$  vertex)
\beq
\label{gaugefermion}
\int d^3x \, \Tr \Big[ \bar{\psi}_{(1)}^{\hat I}  \g^\mu  A_{(1) \mu} \psi_{(1) \hat I}  - \bar{\psi}_{(1)}^{\hat I} \g^\mu \psi_{(1)  \hat I} {A}_{(2) \mu}    \Big]
\eeq 
\noindent 3) Yukawa couplings. From the action in \cite{Imamura:2008dt} suitably rotated to Euclidean space we read (from $(-S)$ and only terms relevant for our calculation)
\begin{align}
\label{yukawa}
&\frac{2 i}{k} \textrm{Tr}\bigg[ -\e_{AB} \e_{\hat{C}\hat{D}} \bar{\psi}_{(0)}^{\a B} q_{(0)}^{\hat{D}} q_{(1)}^{A} \bar{\psi}_{\a(1)}^{\hat{C}} - \e^{AB} \e^{\hat{C}\hat{D}} \bar{q}_{(0)\hat{C}} \psi_{(0)A}^{\a}  \psi_{\a(1)\hat{D}} \bar{q}_{(1)B} 
\non \\
& \qquad \quad + \bar{\psi}_{(0)}^{\a A} q_{(0)}^{\hat{B}} \psi_{(1)\a\hat{B}} \bar{q}_{(1)A} +\bar{q}_{(0)\hat{B}} \psi_{(0)A}^{\a} q^A_{(1)} \bar{\psi}_{\a(1)}^{\hat{B}} \bigg]   \non \\
& +\frac{2 i}{k}  \textrm{Tr}\bigg[\frac{1}{2} \psi_{(1) \hat{1}}^\a \bar{\psi}_{\a(1)}^{\hat{1}}\bar{q}_{(0) \hat{K}}(\sigma_3)^{\hat{K}}_{\,\,\hat{L}} \,q_{(0)}^{\hat{L}} 
- \frac{1}{2} \psi_{(1) \hat{2}}^\a \bar{\psi}_{\a(1)}^{\hat{2}}\bar{q}_{(0) \hat{K}}(\sigma_3)^{\hat{K}}_{\,\,\hat{L}} \,q_{(0)}^{\hat{L}} 
\non \\
&\qquad \qquad + \psi_{(1) \hat{1}}^\a \bar{\psi}_{\a(1)}^{\hat{2}}\bar{q}_{(0) \hat{2}} q_{(0)}^{\hat{1}} + \psi_{(1) \hat{2}}^\a \bar{\psi}_{\a(1)}^{\hat{1}}\bar{q}_{(0) \hat{1}} q_{(0)}^{\hat{2}} \bigg]
\end{align}

\vskip 10pt
\noindent
Finally, we recall our color conventions. We work with hermitian generators for $U(N_A)$ gauge groups ($\small{A}=0,1,2$), satisfying 
\bea
{\Tr} (T^a_{(A)} T^b_{(A)} ) = \d^{ab}  \quad , \quad
\sum_{a=1}^{N^2_A} (T^a_{(A)})_{ij} (T^a_{(A)})_{kl} = \d_{il} \d_{jk}  \quad , \quad  f^{abc}_{(A)} f^{abc}_{(A)} = 2 N^3_A   
\\
\non
\eea

\section{Useful identities on the unit circle}\label{sec:formulitas}

We parametrize a point on the unit circle $\G$ as 
\beq
x_i^\mu = (\cos{\tau_i}, \sin{\tau_i},0) \quad , \quad \dot{x}_i^\mu = (-\sin{\tau_i}, \cos{\tau_i},0) \quad , \quad | x_i|^2 = 1
\eeq 
Simple identities that turn out to be useful along the calculation are
\bea
\label{I1}
&& (x_i - x_j)^2 = 4 \sin^2{\frac{\tau_i-\tau_j}{2}}
\\
\label{I2}
&& x_i \cdot x_j = \dot{x}_i \cdot \dot{x}_j = \cos{(\tau_i - \tau_j)}
\\
\label{I3}
&& x_i \cdot \dot{x}_j = \sin{(\tau_i - \tau_j)}
\\
\label{I4}
&& (x_i \cdot x_j) (\dot{x}_i \cdot \dot{x}_j)  - (x_i \cdot \dot{x}_j ) (\dot{x}_i \cdot x_j ) = 1
\\
\label{I5}
&& (x_i-x_j) \cdot  (\dot{x}_i + \dot{x}_j) = 2 \sin{(\tau_i - \tau_j)}
\eea

We now consider bilinears constructed in terms of $c$ spinors in \cite{Cooke:2015ila}. These are different for the two kinds of femionic WL. 

\vskip 10pt
\noindent
\underline{The $\psi_1$-loop}: In this case we have
\bea \label{c1}
&& c(\tau) = \frac{C}{\cos{\frac{\tau}{2}} + \sin{\frac{\tau}{2}}} (\cos{\tau}, 1 + \sin{\tau}) = C ( \cos{\tfrac{\tau}{2}} - \sin{\tfrac{\tau}{2}}, \cos{\tfrac{\tau}{2}} + \sin{\tfrac{\tau}{2}})
\non \\
&& \bar{c}(\tau) = \frac{\bar C}{\cos{\frac{\tau}{2}} - \sin{\frac{\tau}{2}} }  \left( \begin{array}{c}  
 1 - \sin{\tau} \\
\cos{\tau}
 \end{array}\right) = \bar C \left( \begin{array}{c}  
\cos{\frac{\tau}{2}} - \sin{\frac{\tau}{2}}  \\
\cos{\frac{\tau}{2}} + \sin{\frac{\tau}{2}}
 \end{array}\right)
 \eea
with $C \bar C = -\tfrac{i}{k}$. Writing $c_i \equiv c(\tau_i)$ we have
\bea
&& (c_{i} \bar c_{j}) = -\frac{2i}{k} \cos{\frac{\tau_i-\tau_j}{2}}
 \label{eq:etaetabar1}   \\
&& (c_{i} \gamma^1 \bar c_{j}) =   \frac{2i}{k} \sin{\frac{\tau_i + \tau_j}{2}}
\label{eq1:gamma1}\\
&& (c_{i} \gamma^2 \bar c_{j}) = -\frac{2i}{k} \cos{\frac{\tau_i + \tau_j}{2}} 
\label{eq1:gamma2}\\
&& (c_{i} \gamma^3 \bar c_{j}) = \frac{2}{k} \sin{\frac{\tau_i-\tau_j}{2}}
\label{eq:etagamma01}  
\\
&& (c_{i} \gamma_{\mu} \bar c_{j})\, (x_i - x_j)^{\mu} =  -\frac{4i}{k} \, \sin{\frac{\tau_i - \tau_j}{2}}  
\label{eq:etagammax1}
\eea
More generally, we can write
\beq
\label{eq:cgammac1}
(c_{i} \gamma^\mu \bar c_{j}) = \frac{2}{k^2} \, \frac{1}{(c_{i} \bar c_{j})} \Big[ -\dot{x}_i^{\mu} - \dot{x}_j^{\mu} + i \, \varepsilon^{\mu\nu\rho}  \, \dot{x}_i^\nu \, \dot{x}_j^\rho \Big]
\eeq

\vskip 10pt
\noindent
\underline{The $\psi_2$-loop}: In this case we have
\bea\label{c2}
&& d(\tau) = \frac{D}{\cos{\frac{\tau}{2}} - \sin{\frac{\tau}{2}}} (-\cos{\tau}, 1 - \sin{\tau}) = - D (\cos{\tfrac{\tau}{2}} + \sin{\tfrac{\tau}{2}}, -\cos{\tfrac{\tau}{2}} + \sin{\tfrac{\tau}{2}})
\non \\
&& \bar{d}(\tau) = \frac{\bar D}{\cos{\frac{\tau}{2}} + \sin{\frac{\tau}{2}} }  \left( \begin{array}{c}  
 1 + \sin{\tau} \\
-\cos{\tau}
 \end{array}\right) = \bar D \left( \begin{array}{c}  
 \cos{\frac{\tau}{2}} + \sin{\frac{\tau}{2}} \\
-\cos{\frac{\tau}{2}} + \sin{\frac{\tau}{2}}
 \end{array}\right)
 \eea
with $D \bar D = \tfrac{i}{k}$, and the corresponding bilinears are
\bea
&& (d_{i} \bar d_{j}) = -\frac{2i}{k} \cos{\frac{\tau_i-\tau_j}{2}}
 \label{eq:etaetabar2}   \\
&& (d_{i} \gamma^1 \bar d_{j}) =   -\frac{2i}{k} \sin{\frac{\tau_i + \tau_j}{2}}
\label{eq2:gamma1}\\
&& (d_{i} \gamma^2 \bar d_{j}) =  \frac{2i}{k} \cos{\frac{\tau_i + \tau_j}{2}} 
\label{eq2:gamma2}\\
&& (d_{i} \gamma^3 \bar d_{j}) = \frac{2}{k} \sin{\frac{\tau_i-\tau_j}{2}}
\label{eq:etagamma02}  
\\
&& (d_{i} \gamma_{\mu} \bar d_{j})\, (x_i - x_j)^{\mu} =  \frac{4i}{k} \, \sin{\frac{\tau_i - \tau_j}{2}}  
\label{eq:etagammax2}
\eea
More generally, we can write
\beq
\label{eq:cgammac2}
(d_{i} \gamma^\mu \bar d_{j}) = \frac{2}{k^2} \frac{1}{(c_{i} \bar c_{j})} \Big[ \dot{x}_i^{\mu} + \dot{x}_j^{\mu} + i \, \varepsilon^{\mu\nu \rho} \, \dot{x}_i^\nu \, \dot{x}_j^\rho \Big]
\eeq
We note a sign difference in the $\mu=1,2$ bilinears of the two WL (formulae (\ref{eq1:gamma1}, \ref{eq1:gamma2}) vs. (\ref{eq2:gamma1}, \ref{eq2:gamma2})).  

\section{Parity and reality of a generic WL diagram} \label{sec:alldetails}

Here we prove that for any loop diagram at order $(1/k)^L$ with $n_S$ contour insertions of the scalar bilinears,  the number $n_\g$  of fermion bilinears $(c\g\bar c)$ that get produced after $\g$--algebra reduction has the same parity of $L+n_S$.  This result is crucial to prove identity (\ref{relation}) in the main text. 

To this end, we consider a  diagram containing $n_S$ scalar, $2n_F$ fermion and $n_A$ gauge couplings from the WL expansion (see Fig. \ref{count}). Moreover, we assume that the bulk of the diagram is built up with $i_A$ cubic gauge vertices, $i_S$ esa--scalar vertices, $i_Y$ Yukawa couplings, $i_{AF}$  gauge--fermion vertices, $i_{AS}$ cubic and $j_{AS}$ quartic gauge--scalar vertices, $i_{AG}$ cubic gauge--ghost vertices, and $I_A$ gauge, $I_{G}$ ghost, $I_S$ scalar and  $I_F$ fermion propagators, respectively. These assignments are summarized in Table \ref{power}.

From the structure of the vertices we have the following constraints
\bea 
\label{constraints}
&&  2 I_A = n_A + 3i_A + i_{AF} + i_{AS} + 2 j_{AS} + i_{AG} 
\non \\
&&  I_F =  n_F +  i_{AF} +  i_Y
\non \\
&& I_S = n_S + 3 i_S +  i_Y + i_{AS} + j_{AS}
\non \\
&&  I_G =  i_{AG} 
\eea

\begin{table}[h]
\begin{tabular}{cccccc}
$I_A$ & {\includegraphics[scale=0.26]{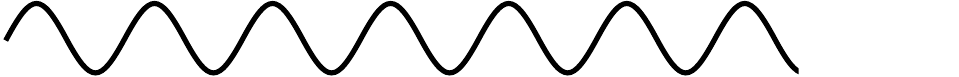}}  & \hspace{1cm} $i_{AF}$ & \hspace{0cm}\raisebox{-0.8cm}{\includegraphics[scale=0.25]{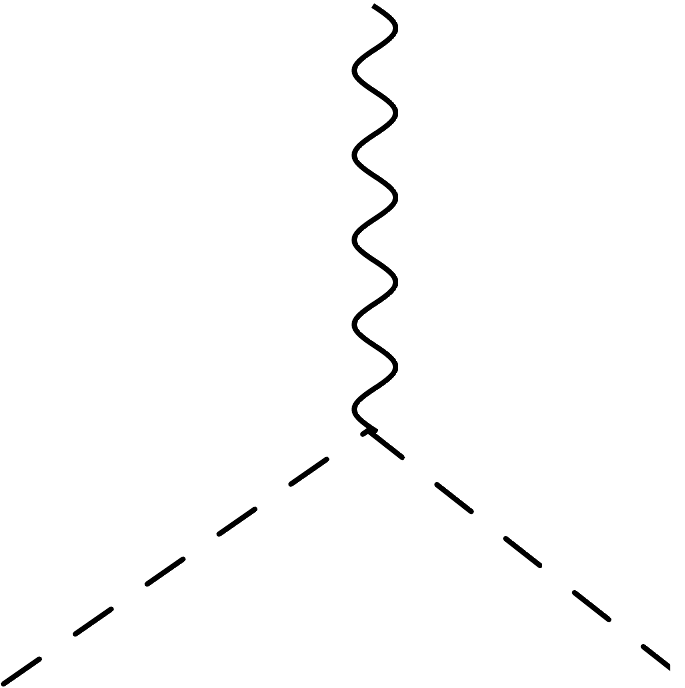}} & \hspace{1.8cm}$i_{A}$ &\hspace{-0.2cm}\raisebox{-0.8cm} {\includegraphics[scale=0.25]{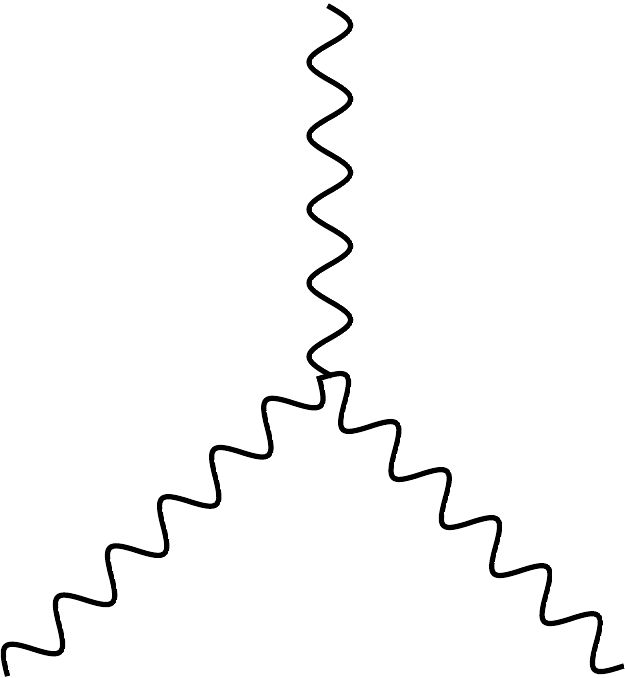}}\vspace{0.4cm}\\

$I_S$ & \raisebox{0.1cm} {\includegraphics[scale=0.26]{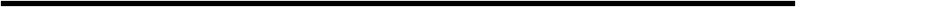}}  & \hspace{1cm} $i_{AS}$ & \hspace{-0.1cm}\raisebox{-0.8cm}{\includegraphics[scale=0.25]{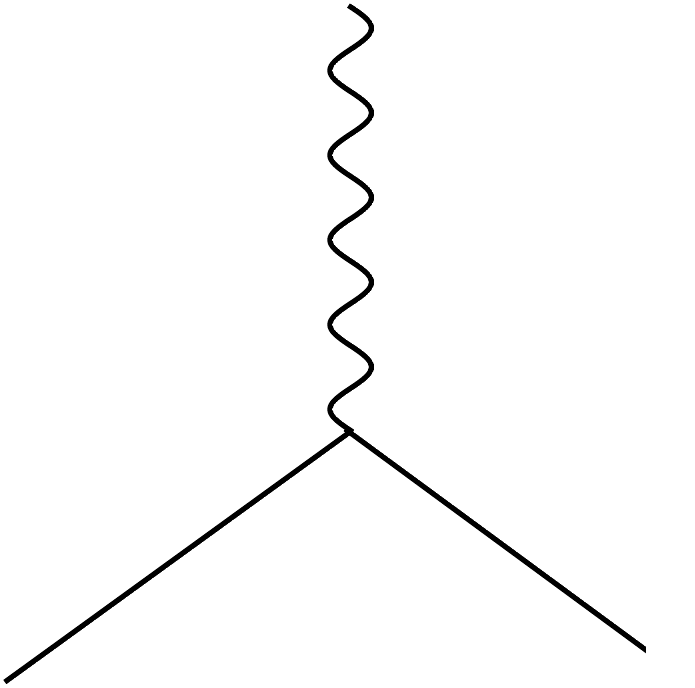}}  & \hspace{1.6cm} $i_{S}$ & \hspace{0.1cm}\raisebox{-0.8cm} {\includegraphics[scale=0.27]{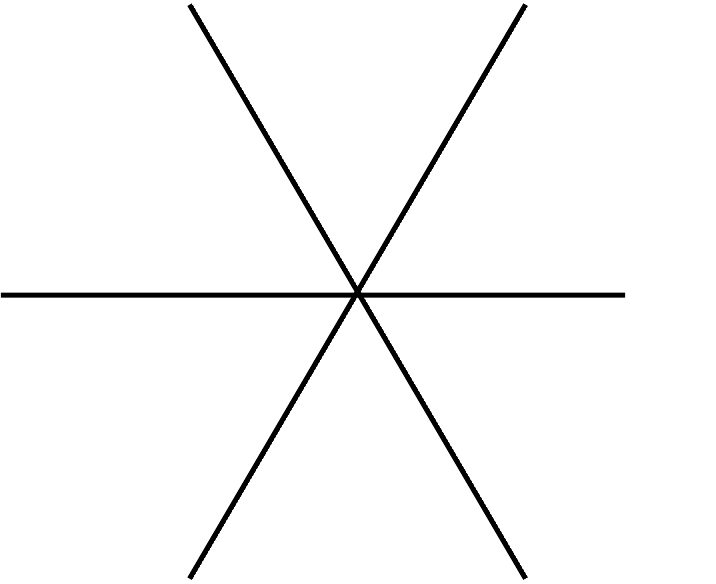}} \vspace{0.4cm} \\

$I_F$ & \!\!\!\!\!\!{\includegraphics[scale=0.25]{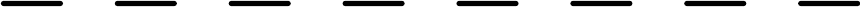}}  &   \hspace{1cm}  $j_{AS}$ & \hspace{0cm}\raisebox{-0.8cm}{\includegraphics[scale=0.26]{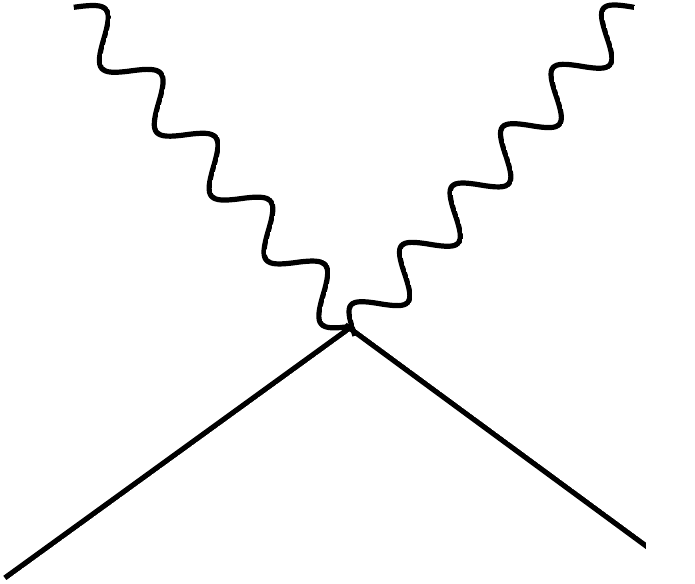}}  & \hspace{1.5cm} $i_{AG}$  &\hspace{-0.25cm} \raisebox{-0.8cm}{\includegraphics[scale=0.25]{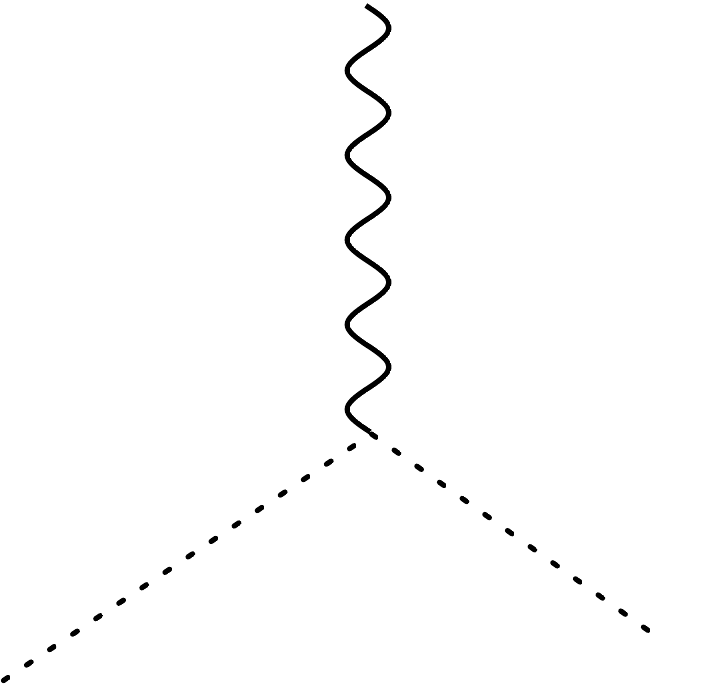}}\vspace{0.4cm} \\

$I_{G}$ & \!\!\!\!\!\!\raisebox{0.1cm} {\includegraphics[scale=0.25]{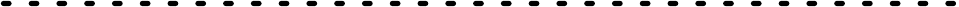}} &  \hspace{1cm}$i_{Y}$ & \hspace{0cm}\raisebox{-0.8cm}{\includegraphics[scale=0.26]{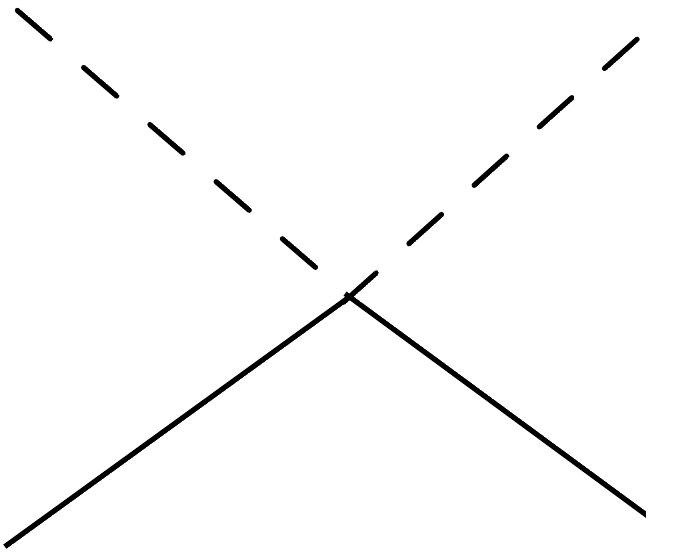}} &  \vspace{0.4cm} 
\end{tabular}
\caption{Definition of number of propagators and vertices.}
\label{power}
\end{table} 

We begin by proving the following  statement
\beq
\label{parity}
L + n_S = [(i_Y + n_F) + (I_A + i_A)]  \, {\rm mod}(2) =  [n + n_\varepsilon ]  \, {\rm mod}(2)
\eeq
where $n$ is the total number of initial gamma matrices (coming from fermionic propagators and $i_{AF}$ vertices) distributed in $n_F$ bilinears, and $n_\varepsilon$ is the total number of initial epsilon tensors (coming from gauge propagators and cubic gauge vertices). 

Now, taking into account the Feynman rules in Appendix \ref{sec:conventions} the power $L$ in the coupling constant $1/k$ is given by
\bea
\label{L}
L &=& n_F + n_S + I_A - i_A + i_Y + 2i_S + I_G - i_{AG}  
\non \\
&=& n_F + n_S + I_A - i_A + i_Y + 2i_S 
\eea
where the last identity in  (\ref{constraints}) has been used.  

Moreover, the number $n$ of original gamma matrices (coming from fermion propagators and $i_{AF}$ vertices) and the number $n_{\varepsilon}$ of original $\varepsilon$ tensors (coming from gauge propagators and $i_A$ vertices) are
\bea
\label{n}
&& n =\# ~ {\text {gamma matrices}} = I_F + i_{AF} = n_F + i_Y + 2i_{AF}
 \non \\
 && n_{\varepsilon} = \# ~\varepsilon ~{\text{tensors}} = I_A + i_A
 \eea
where the second identity in (\ref{constraints}) has been used.  Merging results (\ref{L}) and (\ref{n}) we finally obtain identity 
(\ref{parity}) that allows us to trade the parity of  $L + n_S$ with that of $n + n_{\varepsilon}$. \\

We then study the two cases, $L+n_S$ even or odd, by separately discussing the four possible configurations
\bea
\begin{matrix}
(L+n_S) \\
{\text {even}} 
\end{matrix}
\Rightarrow
~\Big\{
\begin{matrix}
~{\text {(1a)}}~ (n,  n_\varepsilon) = {\text{(even, even)}} \\
{\text {(1b)}} ~(n,  n_\varepsilon) = {\text{(odd, odd)}} 
\end{matrix}
\qquad
\begin{matrix}
(L+n_S) \\
{\text {odd}} 
\end{matrix}
\Rightarrow
~  \Big\{
\begin{matrix}
{\text {(2a)}}~ (n,  n_\varepsilon) = {\text{(even, odd)}} \\
{\text {(2b)}} ~(n,  n_\varepsilon) = {\text{(odd, even)}} 
\end{matrix}
\non
\eea
and prove that in the first two configurations $n_\g$ turns out to be even, whereas in the last two ones it is odd. 

In case (1a), the condition that the total number of gamma matrices $n$ must be even implies that the matrices can be distributed among an arbitrary (but $\leq n_F$) number of bilinears containing an even number of matrices times an even number of bilinears containing an odd number of matrices. Therefore, taking into account  reductions (\ref{even}, \ref{odd}) that follow from gamma matrix identities, the initial structure of the contribution from this diagram can be sketchily written as 
\bea
\label{sketchy}
{\rm (even \, \# \, of \, \varepsilon ) } \times \underbrace{[(c {\bar c}) + \varepsilon (c \g {\bar c}) ] \cdots [(c {\bar c}) + \varepsilon (c \g {\bar c}) ] }_{\text{any }\# \leq n_F} 
\times \underbrace{[\varepsilon(c {\bar c}) + (c \g {\bar c}) ] \cdots [\varepsilon(c {\bar c}) + (c \g {\bar c}) ]}_{\text {even \#}} 
\non \\
\eea
After performing all the products, the planarity of the contour implies that non--vanishing contributions will arise only from terms containing an even total number of 
epsilon tensors. In fact, any string of an odd number of tensors can be always reduced to a linear combination of products of Kronecker deltas times one epsilon tensor that would be necessarily contracted with external indices.  

Therefore, in the product of the square brackets in (\ref{sketchy}) we can have an even number of $\varepsilon (c \g {\bar c})$ from the first set of brackets times an even number of $\varepsilon (c {\bar c})$ from the second set. But since the total number of second type of brackets is even, this implies having an even number of $(c \g {\bar c})$ as well. Therefore,  the only non--vanishing products will contain a total number $n_\g$ of $(c \g {\bar c})$ bilinears which is even. 
Otherwise, we can have an odd number of $\varepsilon (c \g {\bar c})$ from the first set of brackets times an odd number of $\varepsilon (c {\bar c})$ from the second set. But since the total number of second type of brackets is even, this implies having an odd number of $(c \g {\bar c})$ from the second set. Therefore, this leads still to a total number $n_\g$ which is (odd + odd) = even.

Let's consider case (1b). Since the number $n$ of gamma matrices is odd, this time we have an odd number of bilinears containing an odd number of matrices. The sketchy structure of the result is
\bea
\label{sketchy2}
{\rm (odd \, \# \, of \, \varepsilon ) } \times \underbrace{[(c {\bar c}) + \varepsilon (c \g {\bar c}) ] \cdots [(c {\bar c}) + \varepsilon (c \g {\bar c}) ] }_{\text{any }\# \leq n_F} 
\times \underbrace{[\varepsilon(c {\bar c}) + (c \g {\bar c}) ] \cdots [\varepsilon(c {\bar c}) + (c \g {\bar c}) ]}_{\text {odd \#}} 
\non \\
\eea
Again, performing all the products, the only non--vanishing contributions come from strings containing a total even number of epsilon tensors. This requires having an even number of $\varepsilon (c \g {\bar c})$ from the first set of brackets times an odd number of $\varepsilon (c {\bar c})$ from the second set. But since the total number of second type of brackets is odd, this also implies having an even number of $(c \g {\bar c})$. In conclusion, the only non--vanishing products will contain a total number $n_\g$ of $(c \g {\bar c})$ bilinears which is even. Alternatively, we can have an odd number of $\varepsilon (c \g {\bar c})$ from the first set of brackets times an even number of $\varepsilon (c {\bar c})$ from the second one, which implies having an odd number of $(c \g {\bar c})$. In total, we still end up with an even number $n_\g$. 

Therefore we have proved that for $L+n_S$ even, planarity implies $n_\g$ even.

A similar analysis can be applied to the case where $L +n_S$ is odd. For instance, if we consider (2a) case, the general structure of the contribution reads
\bea
\label{sketchy3}
{\rm (odd \, \# \, of \, \varepsilon ) } \times \underbrace{[(c {\bar c}) + \varepsilon (c \g {\bar c}) ] \cdots [(c {\bar c}) + \varepsilon (c \g {\bar c}) ] }_{\text{any }\# \leq n_F} 
\times \underbrace{[\varepsilon(c {\bar c}) + (c \g {\bar c}) ] \cdots [\varepsilon(c {\bar c}) + (c \g {\bar c}) ]}_{\text {even \#}} 
\non \\
\eea
In order to realize a string containing an overall even number of epsilon tensors, we can take an even number of $\varepsilon(c \g {\bar c})$ from the first set of brackets times an odd number of $\varepsilon(c {\bar c})$ from the second one. But since the number of brackets in the second set is even, this implies having an odd number of $(c \g {\bar c})$ as well.  In total we have (even + odd) number of $(c \g {\bar c})$ bilinears, leading to $n_\g$ odd. The same conclusion is reached if we alternatively take an odd number of $\varepsilon(c \g {\bar c})$ from the first set of brackets times an even number of $\varepsilon(c {\bar c})$ from the second one that comes together with an even number of 
$(c \g {\bar c})$.

The analysis of case (2b) goes similarly and we are led to the conclusion that for $L+n_S$ odd, planarity implies $n_\g$ odd. We have then proved that $n_\g$ always has the same parity of $L+n_S$. 

\vskip 15pt

We conclude this Appendix with an analysis of the reality of the perturbative expansion of fermionic WL. We will prove that  the result at any order is always real, as a consequence of the planarity of the contour and the fact that we work at framing zero. 

In order to prove it, we apply counting arguments similar to the ones used above, this time keeping track of the different sources of the immaginary unit $i$. 

Focusing on $W_{\psi_1}$ in (\ref{WL1}) we first notice that from expansion of the Wilson loop we have a factor $i^{n_A+2 n_F}$. Moreover, as explained in Section \ref{sec:relation} each fermionic bilinear can be always reduced to a linear combination of expressions (\ref{eq:etaetabar1}-\ref{eq:etagamma01}). However, the planarity of the contour eventually rules out the appearance of $\g^3$ bilinear. Since all the other ones contain an $i$ factor, we can count an additional immaginary unit for each of the $n_F$ structures. We are thus  left with an overall power $i^{(n_A+n_F) \, ({\rm mod} ~2)}$. Next we count the $i$ factors coming from internal vertices and propagators, getting a further power $i^{I_F+ I_A+i_{AS}+ i_Y+i_{AG}}$. Putting everything together we are left with a total power $i^p$ with
\begin{align}
p & = n_A+n_F+I_F+ I_A+i_{AS}+ i_Y+i_{AG}  \quad \textrm{(mod 2)} 
\end{align}
Making repeated use of identities (\ref{constraints}) this can be rewritten as
\beq
 p =  I_A+i_A \quad\textrm{(mod 2)} 
\eeq
But, as discussed above,  $I_A+i_A=n_{\e}$, which is the number of initial epsilon tensors. Therefore we have an overall $i^{n_{\e}}$.  Any other $\e$ tensor coming from $\gamma$-algebra reduction always enters with an additional $i$ (see identities in Appendix \ref{sec:conventions}). We thus have a total factor $(i \e)^{n_{\e}+m}$ and, from planarity and at framing zero, we must have $n_{\e}+m={\rm even}$. Therefore, we end up with an even number of  $i$ and the result is always real, independently of the pertubative order. 
Thanks to identity (\ref{relation}) this result extends trivially to $W_{\psi_2}$.

 \section{Useful formulae for the matrix model analysis}
 \label{appendixF}
 The expression for $B_4(\Lambda_A)$ and $ C_4(\Lambda_A,\Lambda_{A+1})$ appearing in the expansion of $Q_A$ are given by
  \begin{align}
 B_4(\Lambda_A)=&\frac{1}{90} \biggl(\left(5 N_A^2-3\right) \text{Tr}\left(\Lambda_A^2\right)^2-N_A \text{Tr}\left(\Lambda_A ^4\right)-10 N_A \text{Tr}\left(\Lambda_A ^2\right) \text{Tr}(\Lambda_A )^2+\nonumber\\
 &+4 \text{Tr}\left(\Lambda_A ^3\right) \text{Tr}(\Lambda_A )+5\text{Tr}(\Lambda_A )^4\biggr)\\
 C_4(\Lambda_A,\Lambda_{A+1})=&\frac{1}{24 } \biggl (3 N_{A+1}^2 \text{Tr}\left(\Lambda_A ^2\right)^2+6
   \text{Tr}\left(\Lambda_{A+1}^2\right) \left((N_{A+1} N_{A}-2) \text{Tr}\left(\Lambda_A^2\right)\right.-\nonumber\\
    &\left.-2 N_{A} \text{Tr}(\Lambda_A ) \text{Tr}(\Lambda_{A+1})\right)-\left. 2 N_{A+1}
   \text{Tr}\left(\Lambda_A ^4\right)-\right.\nonumber\\
   &\left.-12 N_{A+1} \text{Tr}(\Lambda_A )
   \text{Tr}\left(\Lambda_A ^2\right) \text{Tr}(\Lambda_{A+1})+3 N_{A}^2
   \text{Tr}\left(\Lambda_{A+1}^2\right)^2-\right.\nonumber\\ 
   &-2 N_{A} \text{Tr}\left(\Lambda_{A+1}^4\right)+8
   \text{Tr}(\Lambda_A ) \text{Tr}\left(\Lambda_{A+1}^3\right)+8
   \text{Tr}\left(\Lambda_A ^3\right) \text{Tr}(\Lambda_{A+1})+\nonumber\\
   &+12
   \text{Tr}(\Lambda_A )^2 \text{Tr}(\Lambda_{A+1})^2\biggr).
 \end{align}
Consider now the  gaussian model defined by the matrix integral
\be
\int d \Lambda~ e^{- \alpha \mathrm{Tr}(\Lambda^2)}
\ee
The expectation values that we have used in our analysis are 
\be
\begin{split}
\langle\mathrm{Tr}(\Lambda^{2 k})\rangle_0= 
\alpha^{-k}\frac{(2 k)! }{\left(2^k
   k!\right) } \sum _{j=0}^k \binom{k}{j} \binom{N}{k-j+1} 2^{-j}
   \end{split}
\ee
and
\be
\begin{split}
\langle{\mathrm{Tr}(\Lambda^2)^m\mathrm{Tr}(\Lambda)^{2 k}}\rangle_0=&\left. \frac{(-1)^m}{\int d \Lambda~ e^{- \alpha\mathrm{Tr}(\Lambda^2)}}\frac{d^m}{d\alpha^m} 
\frac{d^{2 k}}{d y^{2 k}} \int d \Lambda~ e^{- \alpha \mathrm{Tr}(\Lambda^2)+y \mathrm{Tr}(\Lambda) } 
\right|_{y=0}=\\
=&\left. \frac{(-1)^m}{\int d \Lambda~ e^{- \alpha\mathrm{Tr}(\Lambda^2)}}\frac{d^m}{d\alpha^m}
\frac{d^{2 k}}{d y^{2 k}}\left( {\int d \Lambda~ e^{- \alpha \mathrm{Tr}(\Lambda^2)+y \mathrm{Tr}(\Lambda) } }{}\right)\right|_{y=0}=\\=&\left.
\frac{(-1)^m}{\int d \Lambda~ e^{- \alpha\mathrm{Tr}(\Lambda^2)}} \frac{d^m}{d\alpha^m}
\frac{d^{2 k}}{d y^{2 k}}\left( e^{\frac{N y^2}{4\alpha}}{\int d \Lambda~ e^{- \alpha \mathrm{Tr}[(\Lambda)^2]} }{}\right)\right|_{y=0}=\\
&=
\left.
(-1)^m \left(\frac{\pi}{\alpha}\right)^{-\frac{N^2}{2}}\frac{d^m}{d\alpha^m}
\frac{d^{2 k}}{d y^{2 k}}\left( e^{\frac{N y^2}{4\alpha}} \left(\frac{\pi}{\alpha}\right)^{\frac{N^2}{2}}\right)\right|_{y=0}
\end{split}
\ee

\section{Cancellation of gauge dependent terms}\label{gaugedependent}
In the computation of diagrams (a), (c) and (e) we have neglected the contributions from one-loop corrected gauge propagator (\ref{1vector}) containing the double derivatives. As already mentioned in Sec. \ref{lifting}, we expect these gauge dependent contributions to cancel each others. Here we confirm this expectation.

The gauge dependent contribution from diagram (a) reads
\begin{align} 
 {\rm (a)}_{g} = &   - \frac{C_{\rm ab}}{4\e (1+2 \e)} \!\int\! d\tau_{1>2>3} \bigg[ (c_3 \g_{\mu} \g_{\nu}\g_{\rho} \bar{c}_2) \dot{x}_{1\eta}  \partial_2^{\rho} \partial_3^{\mu} \partial_1^{\eta}\partial_1^{\nu}  \int \! d^{3-2\e} w \frac{(x_{1w}^2)^{2\e}}{(x_{2w}^2)^{1/2-\e}}\frac{1}{(x_{3w}^2)^{1/2-\e}} \non \\ & \hspace{4cm}   - \big(1\!\rightarrow\!2\!\rightarrow\!3\!\rightarrow\!1\big) +\big(3\!\rightarrow\!2\!\rightarrow\!1\!\rightarrow\!3\big)\,\, \bigg] 
\end{align}
with $C_{\rm ab}$ defined in (\ref{CAB}).  Working out the $\g$-algebra and performing the integrations we obtain
\beq \label{gauge1}
{\rm (a)}_g = \frac{N_0 N_1^2 N_2 }{(N_1+N_2)k^3} \frac{e^{3 \g_E \e}}{4^4 \p^{1-3\e}} \,\, 48
\eeq
The gauge dependent part of diagram (c) produces a correction to the fermion propagator of the form 
\begin{align}
&\frac{N_0 N_1}{k^2} \textrm{Tr}(\bar{\psi}(p)\g^\m\psi(-p)) \frac{p_\m}{(p^2)^{2\e}} I_{{\rm (c)}_g} 
\end{align}
with
\begin{align}
I_{{\rm (c)}_g}&  =- \frac{\csc(2\e\p)\sec(\e\p)\G(3/2-\e)}{2^{5-6\e}\p^{1/2-2\e}\G(3/2-3\e)\G(1-\e)\G(3/2+\e)}=-\frac{1}{32\p^2 \e} + \frac{-1+\g_E-\log(4\p)}{16\p^2} 
\end{align}
This can be inserted into the loop contour to get
\begin{align}\label{gauge2}
 {\rm (c)}_g = - \frac{N_0 N_1^2 N_2 }{(N_1+N_2)k^3} \frac{e^{3 \g_E \e}}{4^4 \p^{1-3\e}} \,\, 24
\end{align}
The gauge dependent part coming from diagram (e) is given by
\begin{align}\label{gauge3}
{\rm (e)}_g= &\,\, -\frac{ C_{\rm ef}}{1+2\e}  \int d\tau_{1>2>3>4} \,\bigg[  \bigg( \sin^2 \frac{\t_{12}}{2} \bigg)^{-1+\e} \frac{ (4 \e \cos^2 \frac{\t_{34}}{2}-1)}{\big(\sin^2 \frac{\t_{34}}{2}\big)^{1-2\e}} + {\rm cyclic}  \bigg]  \non  
\end{align} 
 where $C_{\rm ef}$  and ``cyclic'' are defined in (\ref{CEF}) and below. Solving the integral we get 
 \begin{align}
 {\rm (e)}_g& = - \frac{N_0 N_1^2 N_2 }{(N_1+N_2)k^3} \frac{e^{3 \g_E \e}}{4^4 \p^{1-3\e}} \,\, 24
\end{align}
It is immediate to see that  $(\ref{gauge1})+(\ref{gauge2})+(\ref{gauge3})=0$.

\section{Details on diagrams (a) and (b)} \label{sec:ab}
 
Here we give details on the calculation of the two integrals appearing in eqs. (\ref{apar}, \ref{bpar})
\begin{align} 
{\rm (a)}_{\psi_1} & =  C_{\rm ab} \!\int\! d\tau_{1>2>3} \bigg[ (c_3 \g_{\mu} \g_{\nu}\g_{\rho} \bar{c}_2) \dot{x}_1^{\nu}  \partial_2^{\rho} \partial_3^{\mu} \, \textrm{I(2,1,1)} - \big(1\!\rightarrow\!2\!\rightarrow\!3\!\rightarrow\!1\big) +\big(3\!\rightarrow\!2\!\rightarrow\!1\!\rightarrow\!3\big) \bigg] \label{apar2} \\
{\rm (b)}_{\psi_1}  & =  - C_{\rm ab} \int \!d\tau_{1>2>3}  \bigg[ (c_3 \g_{\mu}\g_{\rho} \bar{c}_2) \partial_2^{\rho} \partial_3^{\mu} \, \textrm{I(2,1,1)} - \big(1\!\rightarrow\!2\!\rightarrow\!3\!\rightarrow\!1\big) +\big(3\!\rightarrow\!2\!\rightarrow\!1\!\rightarrow\!3\big) \bigg] \label{bpar2}
\end{align}
with $I(2,1,1)$ defined in (\ref{integral}). In both cases we focus on  the first contribution, while adding the cyclic permutations later on. We are eventually interested in the result $[{\rm (a) + (b)}]$.

One possibile way to get rid of the derivatives is to first Feynman parametrize $ \textrm{I(2,1,1)} $ and integrate over the internal point $w$. From
\beq
\textrm{I(2,1,1)}  = \frac{\G(\frac{1}{2}-3\e)\p^{3/2-\e}}{\G(\frac{1}{2}-\e)^2\G(1-2\e)} \int [d\a]_3 \frac{\a_1^{-2\e}(\a_2\a_3)^{-1/2-\e}}{ \big(\a_1\a_2 x_{12}^2+\a_2\a_3 x_{23}^2+\a_1\a_3 x_{13}^2\big)^{1/2-3\e}}
\eeq 
we obtain
 \begin{align} \label{deriv}
 \partial_2^{\rho} \partial_3^{\mu} \textrm{I(2,1,1)}  & =  
\frac{\G(\frac{5}{2}-3\e)\p^{3/2-\e}}{\G(\frac{1}{2}-\e)^2\G(1-2\e)} \int [d\a]_3 \frac{ 4 \a_1^{-2\e}(\a_2\a_3)^{-1/2-\e}}{ \big(\a_1\a_2 x_{12}^2+\a_2\a_3 x_{23}^2+\a_1\a_3 x_{13}^2\big)^{5/2-3\e}} \, \times \non \\
&\qquad  \bigg(\a_1\a_2^2\a_3 x_{12}^{\rho}x_{23}^{\mu}+\a_1^2\a_2\a_3 x_{12}^{\rho}x_{13}^{\mu}-\a_2^2\a_3^2 x_{23}^{\rho}x_{23}^{\mu}-\a_1\a_2\a_3^2 x_{23}^{\rho}x_{13}^{\mu}\bigg)
\non \\
& +  \frac{\G(\frac{3}{2}-3\e)\p^{3/2-\e}}{\G(\frac{1}{2}-\e)^2\G(1-2\e)} \int [d\a]_3 \frac{2\a_1^{-2\e}(\a_2\a_3)^{1/2-\e} \,\hat{\eta}^{\rho\mu}}{ \big(\a_1\a_2 x_{12}^2+\a_2\a_3 x_{23}^2+\a_1\a_3 x_{13}^2\big)^{3/2-3\e}} 
\end{align}
We begin by analyzing the first integral in (\ref{deriv}), once inserted in (\ref{apar2}) and (\ref{bpar2}). We need to work out the following bilinears for diagram (a)
\begin{align}
(c_3 \g_{\mu} \g_{\nu}\g_{\rho} \bar{c}_2)\dot{x}_1^{\nu}  x_{12}^{\rho}x_{23}^{\mu} & = -\frac{4 i}{k} \sin(\t_{12})\sin\left(\frac{\t_{23}}{2}\right) \\
(c_3 \g_{\mu} \g_{\nu}\g_{\rho} \bar{c}_2)\dot{x}_1^{\nu}  x_{12}^{\rho}x_{13}^{\mu} & = -\frac{8 i}{k} \sin\left(\frac{\t_{12}}{2}\right)\sin\left(\frac{\t_{13}}{2}\right) \\
(c_3 \g_{\mu} \g_{\nu}\g_{\rho} \bar{c}_2)\dot{x}_1^{\nu}  x_{23}^{\rho}x_{23}^{\mu} & = -\frac{8 i}{k} \cos\left(\frac{\t_{12}+\t_{13}}{2}\right)\sin^2\left(\frac{\t_{23}}{2}\right) \label{d1} \\
(c_3 \g_{\mu} \g_{\nu}\g_{\rho} \bar{c}_2)\dot{x}_1^{\nu}  x_{23}^{\rho}x_{13}^{\mu} & = -\frac{4 i}{k} \sin\left(\t_{13} \right)\sin\left(\frac{\t_{23}}{2}\right)
\end{align}
and the corresponding ones for diagram (b)
\begin{align}
(c_3 \g_{\mu}\g_{\rho} \bar{c}_2) x_{
12}^{\rho}x_{23}^{\mu} & = -\frac{4 i}{k} \sin\left(\t_1-\t_2 \right)\sin\left(\frac{\t_{23}}{2}\right) \\
(c_3 \g_{\mu}\g_{\rho} \bar{c}_2)  x_{12}^{\rho}x_{13}^{\mu} & = -\frac{8 i}{k} \sin\left(\frac{\t_{12}}{2}\right)\sin\left(\frac{\t_{13}}{2}\right)\\
(c_3 \g_{\mu}\g_{\rho} \bar{c}_2) x_{23}^{\rho}x_{23}^{\mu} & = -\frac{4 i}{k} \cos\left(\frac{\t_{23}}{2}\right)\bigg(1-\cos(\t_{23})\bigg) \label{d2} \\
(c_3 \g_{\mu}\g_{\rho} \bar{c}_2)  x_{23}^{\rho}x_{13}^{\mu} & = -\frac{4 i}{k} \sin\left(\t_{13} \right)\sin\left(\frac{\t_{23}}{2}\right) 
\end{align} 
It is easy to see that if we consider the sum  $[{\rm (a) + (b)}]$, most of the bilinear terms cancel and we are only left with the difference between (\ref{d1}) and (\ref{d2}). Inserting the result into the integrals and restoring the cyclic permutations we find
\begin{align}
& - i\frac{C_{\rm ab} }{k}\frac{\G(\frac{5}{2}-3\e) 4^3\p^{3/2-\e}}{\G(\frac{1}{2}-\e)^2\G(1-2\e)} \!\int\! d\tau_{1>2>3} \bigg[ \sin\left(\frac{\t_{12}}{2}\right)\sin\left(\frac{\t_{13}}{2}\right) \sin^2\left(\frac{\t_{23}}{2}\right) \non 
\\ & \int [d\a]_3 \frac{  \a_1^{-2\e}(\a_2\a_3)^{3/2-\e}}{ \big(\a_1\a_2 x_{12}^2+\a_2\a_3 x_{23}^2+\a_1\a_3 x_{13}^2\big)^{5/2-3\e}}  + {\rm cyclic} \bigg]
\end{align}
where $C_{\rm ab}$ has been defined in (\ref{CAB}). 

This integral can be further elaborated  by  using the standard two-fold Mellin-Barnes representation for the denominator obtaining
\begin{align} \label{F14}
& \tilde{C}_{\rm ab} \!  \int_{-i \infty}^{-i \infty} \frac{du dv}{(2\p i)^{2}}\G(-u)\G(-v)\G(u+v+5/2-3\e) \G(2\e-u)\G(2\e-v)\G(u+v+1-2\e)\times  \non \\ &\!\int\! d\tau_{1>2>3} \bigg[ \sin\left(\frac{\t_{12}}{2}\right)^{1+2u}\sin\left(\frac{\t_{13}}{2}\right)^{1+2v} \sin\left(\frac{\t_{23}}{2}\right)^{-3+6\e-2u-2v} + {\rm cyclic} \bigg]
\end{align}
with
\beq
 \tilde{C}_{\rm ab} =\frac{N_0 N_1^2 N_2 }{(N_1+N_2)k^3}\frac{\G(\frac{1}{2}-\e)^2}{\G(1-2\e)\G(1+2\e)}\frac{1}{\p^{9/2-3\e} 2^{6-6\e}} 
\eeq
Exploiting the possibility to perform change of variables in the Mellin--Barnes integrations,
one can prove that  the integrand is symmetric under any exchange of two $\t$'s, although in previous formula this not manifest. Thus we can trade the ordered integration $\int\! d\tau_{1>2>3}$ with a free one $\frac{1}{3!}\int^{2\p}_{0} d\t_1 \int^{2\p}_{0} d\t_2\int^{2\p}_{0} d\t_3$
and use the identity (\ref{J}) of Appendix \ref{sect:nonsym}. We finally obtain  
\begin{align}
&   \tilde{C}_{\rm ab}  \, \int\limits_{-i \infty}^{-i \infty} \frac{du dv}{(2\p i)^{2}}\G(-u)\G(-v)\G(u+v+5/2-3\e) \G(2\e-u)\G(2\e-v)\G(u+v+1-2\e)  \non \\ & \,\,\,\qquad\qquad\qquad \times  4 \pi^{3/2}  \,\frac{\G(1+u)\G(1+v)\G(-1+3\e-u-v)\G(1/2+3\e)}{\G(2+u+v)\G(3\e-v)\G(3\e-u)} 
\end{align}
After expanding in $\e$ the contour integrations can be performed and we obtain the final result
\begin{align} \label{term3}
&2 \pi^{5/2}  \tilde{C}_{\rm ab} \, \bigg[\frac{1}{\e} + \frac{3\g_E - 2(1+6\log2) }{3}\bigg] =\frac{N_0 N_1^2 N_2 }{(N_1+N_2)k^3} \frac{e^{3 \g_E \e}}{4^4 \p^{1-3\e}}\bigg[\frac{8}{\e}-\frac{16}{3}+48\log2\bigg]
\end{align}
A similar approach can be applied to the second integral in (\ref{deriv}). In this case we need the following bilinears  
\begin{align}
\! (c_3 \g_{\mu} \g_{\nu}\g_{\rho} \bar{c}_2)\dot{x}_1^{\nu} \hat{\eta}^{\rho \mu} & = \frac{2 i \,(D-2) }{k} \! \left[ \cos\! \left(\!\frac{\t_{12}}{2}\!\right)\! \cos\!\left(\!\frac{\t_{13}}{2}\!\right)\! -\sin\!\left(\!\frac{\t_{12}}{2}\!\right)\! \sin\!\left(\!\frac{\t_{13}}{2}\!\right)\!  \right] \\
(c_3 \g_{\mu}\g_{\rho} \bar{c}_2) \hat{\eta}^{\rho \mu} & =- \frac{2i D }{k} \cos\left(\frac{\t_{23}}{2}\right) 
\end{align} 
where $D = 3-2\e$. Summing the contributions from diagrams (a) and (b) and inserting back into the integrals we are left with  
\begin{align} \label{billy}
& \frac{C_{\rm ab} \, \G(\frac{3}{2}-3\e) 2 \p^{3/2-\e}}{\G(\frac{1}{2}-\e)^2\G(1-2\e)} \!\int\! d\tau_{1>2>3} \bigg[ \!  \left(- \frac{4i(D-2)}{k} \sin\left(\frac{\t_{12}}{2}\right)\sin\left(\frac{\t_{13}}{2} \right) + \frac{4i(D-1)}{k} \cos\left(\frac{\t_{23}}{2}\right)\! \!     \right) \non
\\ & \int [d\a]_3 \frac{  \a_1^{-2\e}(\a_2\a_3)^{1/2-\e}}{ \big(\a_1\a_2 x_{12}^2+\a_2\a_3 x_{23}^2+\a_1\a_3 x_{13}^2\big)^{3/2-3\e}}  + {\rm cyclic} \bigg]
\end{align}
We evaluate the two different trigonometric structures  in the first line of  (\ref{billy})   separately. 

The first term, after Mellin-Barnes parametrization,  turns out to yield the same trigonometric integral as the one found in (\ref{F14}) and can be elaborated exactly as before
\begin{align} \label{term2}
& \tilde{C}_{\rm ab}\!\left(\tfrac{1}{2}-\e\right)\! \!\!\! \int\displaylimits_{-i \infty}^{-i \infty}\!\!\! \frac{du dv}{(2\p i)^{2}}\G(-u)\G(-v)\G(u+v+3/2-3\e) \G(2\e-u)\G(2\e-v)\G(u+v+1-2\e) \non \\ &\!\int\! d\tau_{1>2>3} \bigg[ \sin\left(\frac{\t_{12}}{2}\right)^{1+2u}\sin\left(\frac{\t_{13}}{2}\right)^{1+2v} \sin\left(\frac{\t_{23}}{2}\right)^{-3+6\e-2u-2v} + {\rm cyclic}  \bigg] \non \\[0.1cm] 
& = \frac{N_0 N_1^2 N_2 }{(N_1+N_2)k^3} \frac{e^{3 \g_E \e}}{4^4 \p^{1-3\e}}\bigg[\frac{8}{\e}-16+48\log2\bigg]
\end{align}
where we have symmetrized the integration region and used identity (\ref{J}).  

The second term in (\ref{billy}), after the introduction of Mellin-Barnes parameters,  produces a slightly different trigonometric structure compared to the previous ones and requires separated treatment.  Its evaluation is  reported in Appendix \ref{sect:nonsym}, while here we use the final result (\ref{nsym}) to obtain
\begin{align}
\label{firstterm}
&  \tilde{C}_{\rm ab} (\e-1)\!\!\!  \int\displaylimits_{-i \infty}^{-i \infty} \!\!\frac{du dv}{(2\p i)^{2}}\G(-u)\G(-v)\G(u+v+3/2-3\e) \G(2\e-u)\G(2\e-v)\G(u+v+1-2\e) \non \\ &\!\!\!\int\! d\tau_{1>2>3} \bigg[ \sin\left(\frac{\t_{12}}{2}\right)^{2u}\! \!\!\sin\left(\frac{\t_{13}}{2}\right)^{2v}\!\!\!  \sin\left(\frac{\t_{23}}{2}\right)^{-3+6\e-2u-2v}\!\! \!\!\cos\left(\frac{\t_{23}}{2}\right)+ {\rm cyclic} \bigg] \non\\[0.1cm] 
&= \frac{N_0 N_1^2 N_2 }{(N_1+N_2)k^3} \frac{e^{3 \g_E \e}}{4^4 \p^{1-3\e}}\,\, \frac{256}{3}
\end{align}
We can now collect all the pieces (\ref{term3}) (\ref{term2}) (\ref{firstterm}) and obtain the final result
\beq
[{\rm (a)+(b)}]_{\psi_1} = \frac{N_0 N_1^2 N_2 }{(N_1+N_2)k^3} \frac{e^{3 \g_E \e}}{4^4 \p^{1-3\e}} \bigg[\frac{16}{\e} + 16 (4+6\log 2) \bigg]
\eeq

\section{Trigonometric integrations} \label{sect:nonsym}

We detail here the evaluation of the  trigonometric integrals of Appendix \ref{sec:ab}. We first need the integrals that enter in  equations (\ref{F14}) and   (\ref{term2}). This type of integrals has been solved in \cite{BGLP2}, where the following general identity was found
\begin{align}
\label{J}
\mathcal{J}(\alpha,\beta,\gamma) = &  \int _{0}^{2\pi}\!\!\!\!d\tau_1 \int _{0}^{2\pi}\!\!\!\!d\tau_2\int _{0}^{2\pi}\!\!\!\! d\tau_3
\left[\sin ^{2}\left(\frac{\tau _{12}}{2}\right)\right]^\alpha
\left[\sin ^{2}\left(\frac{\tau _{23}}{2}\right)\right]^\beta
\left[\sin ^{2}\left(\frac{\tau _{13}}{2}\right)\right]^\gamma
\non \\
& \qquad \quad =8\pi^{3/2} \,
\frac{\Gamma(\tfrac{1}{2}+\alpha) \Gamma(\tfrac{1}{2}+\beta) \Gamma(\tfrac{1}{2}+\gamma) \Gamma(1+\alpha+\beta+\gamma)}
{\Gamma(1+\alpha+\gamma)\Gamma(1+\beta+\gamma)\Gamma(1+\alpha+\beta)}
\end{align}
This identity can be immediately specialized to solve  (\ref{F14}) and   (\ref{term2}). 

Next we concentrate on the non--trivial evaluation of the following general integral  
\begin{align}
\mathcal{I}[\alpha,\beta,\gamma] =  \int d\tau_{1>2>3} & \left[
\left(\sin^2\frac{\tau_{12}}{2}\right)^{\alpha}
\left(\sin^2\frac{\tau_{13}}{2}\right)^{\beta}
\left(\sin^2\frac{\tau_{23}}{2}\right)^{\gamma}
\cos\frac{\tau_{23}}{2}\right.\nonumber\\
&\left. -\left(\sin^2\frac{\tau_{23}}{2}\right)^{\alpha}
\left(\sin^2\frac{\tau_{12}}{2}\right)^{\beta}
\left(\sin^2\frac{\tau_{13}}{2}\right)^{\gamma}
\cos\frac{\tau_{13}}{2}\right.\nonumber\\
&\left. \left(\sin^2\frac{\tau_{13}}{2}\right)^{\alpha}
\left(\sin^2\frac{\tau_{23}}{2}\right)^{\beta}
\left(\sin^2\frac{\tau_{12}}{2}\right)^{\gamma}
\cos\frac{\tau_{12}}{2}
\right]
\end{align}
that enters in equation (\ref{firstterm}). 

After non--trivial change of variables the integral can be put in the simpler form
\begin{equation}
\mathcal{I}[\alpha,\beta,\gamma] =
\pi\int\limits^{2\pi}_{0}d\tau_1\int\limits^{2\pi}_{0}d\tau_2\,
\left(\sin^2\frac{\tau_{1}}{2}\right)^{\alpha}
\left(\sin^2\frac{\tau_{2}}{2}\right)^{\beta}
\left(\sin^2\frac{\tau_{12}}{2}\right)^{\gamma}
\cos\frac{\tau_{12}}{2}
\end{equation}
where one of the contour integrations has been trivially performed.
Up to the $\cos\frac{\tau_{12}}{2}$ factor, this integral is very similar to (\ref{J}). Using the following trigonometric identity
\begin{equation}
2\cos\frac{\tau_{12}}{2}\,=\,
\frac{\sin\frac{\tau_{2}}{2}}{\sin\frac{\tau_{1}}{2}}\,+\,
\frac{\sin\frac{\tau_{1}}{2}}{\sin\frac{\tau_{2}}{2}}\,-\,
\frac{\sin^2\frac{\tau_{12}}{2}}{\sin\frac{\tau_{1}}{2}\sin\frac{\tau_{2}}{2}}
\end{equation}
we can write
\begin{align}
\mathcal{I}[\alpha,\beta,\gamma] &= 
\frac{\pi}{2}\int\limits^{2\pi}_{0}d\tau_1\int\limits^{2\pi}_{0}d\tau_2\,
\left(\sin^2\frac{\tau_{1}}{2}\right)^{\alpha-\tfrac{1}{2}}
\left(\sin^2\frac{\tau_{2}}{2}\right)^{\beta+\tfrac{1}{2}}
\left(\sin^2\frac{\tau_{12}}{2}\right)^{\gamma}\nonumber\\
+&\frac{\pi}{2}\int\limits^{2\pi}_{0}d\tau_1\int\limits^{2\pi}_{0}d\tau_2\,
\left(\sin^2\frac{\tau_{1}}{2}\right)^{\alpha+\tfrac{1}{2}}
\left(\sin^2\frac{\tau_{2}}{2}\right)^{\beta-\tfrac{1}{2}}
\left(\sin^2\frac{\tau_{12}}{2}\right)^{\gamma}\nonumber\\
-&\frac{\pi}{2}\int\limits^{2\pi}_{0}d\tau_1\int\limits^{2\pi}_{0}d\tau_2\,
\left(\sin^2\frac{\tau_{1}}{2}\right)^{\alpha-\tfrac{1}{2}}
\left(\sin^2\frac{\tau_{2}}{2}\right)^{\beta-\tfrac{1}{2}}
\left(\sin^2\frac{\tau_{12}}{2}\right)^{\gamma+1}
\end{align}
Using the expression of the $\mathcal{J}$ integral (\ref{J}) in terms of Gamma functions we finally have
\begin{align}
\mathcal{I}[\alpha,&\beta,\gamma] =
\frac{1}{4}\left[
\mathcal{J}(\alpha-\tfrac{1}{2},\gamma,\beta+\tfrac{1}{2})
+\mathcal{J}(\alpha+\tfrac{1}{2},\gamma,\beta-\tfrac{1}{2})
-\mathcal{J}(\alpha-\tfrac{1}{2},\gamma+1,\beta-\tfrac{1}{2})
\right]
\\
&=2\pi^{3/2}\Gamma(1+\alpha+\beta+\gamma)
\left[\frac{\Gamma(\a)\Gamma(\tfrac{1}{2}+\g)\Gamma(1+\b)}
{\Gamma(\tfrac{1}{2}+\a+\g)\Gamma(\tfrac{3}{2}+\b+\g)\Gamma(1+\a+\b)}\right.
\nonumber\\
&\left. +\frac{\Gamma(1+\a)\Gamma(\tfrac{1}{2}+\g)\Gamma(\b)}
{\Gamma(\tfrac{3}{2}+\a+\g)\Gamma(\tfrac{1}{2}+\b+\g)\Gamma(1+\a+\b)}
-\frac{\Gamma(\a)\Gamma(\tfrac{3}{2}+\g)\Gamma(\b)}
{\Gamma(\tfrac{3}{2}+\a+\g)\Gamma(\tfrac{3}{2}+\b+\g)\Gamma(\a+\b)}
\non 
\right]
\end{align}
which further simplifies to
\begin{equation} \label{nsym}
\mathcal{I}[\alpha,\beta,\gamma]=
4\pi^{3/2}\frac{\Gamma(1+\alpha+\beta+\gamma)\Gamma(1+\a)\Gamma(1+\b)\Gamma(\tfrac{1}{2}+\g)}
{\Gamma(\tfrac{3}{2}+\a+\g)\Gamma(\tfrac{3}{2}+\b+\g)\Gamma(1+\a+\b)}
\end{equation}

\end{document}